\documentclass[aps,pre,showpacs,twocolumn]{revtex4-1}

\pdfoutput=1

\usepackage{amsmath,amsfonts,amssymb,amsbsy,amscd}
\usepackage{ifthen}
\usepackage{hyperref}
\usepackage{graphicx}
\usepackage{color}

\graphicspath{{figs/}}

\definecolor{hreflinkcolor}{rgb}{0.13,0.17,0.83}
\hypersetup{colorlinks=true,urlcolor=hreflinkcolor,
		linkcolor=hreflinkcolor,citecolor=hreflinkcolor}

\newcommand{\beq}{\begin{equation}}
\newcommand{\eeq}{\end{equation}}
\newcommand{\bseq}{\begin{subequations}}
\newcommand{\eseq}{\end{subequations}}

\newcommand{\rf}     [1] {~\cite{#1}}
\newcommand{\refref} [1] {Ref.~\cite{#1}}

\newcommand{\refeq}  [1] {(\ref{#1})}
\newcommand{\refeqs} [2]{(\ref{#1}--\ref{#2})}
\newcommand{\reffig} [1] {Fig.~\ref{#1}}
\newcommand{\refFig} [1] {Figure~\ref{#1}}
\newcommand{\refsect}[1] {Sec.~\ref{#1}}
\newcommand{\refappe}[1] {Appendix~\ref{#1}}

\newcommand{\etal}{{\em et al.}}
\newcommand{\ie}{{i.e.}}

\newcommand{\brk}{\\}

\newcommand{\ii}{{\mathrm i}} 
\DeclareMathOperator{\sgn}{sgn}  
\newcommand{\sump}{\sideset{}{'}\sum}
\renewcommand\Re{\operatorname{Re}}
\renewcommand\Im{\operatorname{Im}}

\newcommand{\ev}[1]{\ensuremath{\mathbf{e}_{#1}}}

\newcommand{\vs}{\ensuremath{v_s}}	
\newcommand{\vv}{\ensuremath{v}}	
\newcommand{\vh}{\ensuremath{u}}	

\newcommand{\Uop}[1]{U(#1)}
\newcommand{\UopI}[1]{U^{-1}(#1)}
\newcommand{\Urep}[5]{\mathrm{U}^{#2#3}{}_{#4#5}(#1)}

\newcommand{\f}{\ensuremath{f}}	
\newcommand{\g}{\ensuremath{g}}	
\newcommand{\E}{\ensuremath{E}}	

\newcommand{\fo}{\ensuremath{\f_0}} 
\newcommand{\Eo}{\ensuremath{\E_0}} 
 
\newcommand{\go}{\ensuremath{\g_0}} 
\newcommand{\goTransl}{\ensuremath{\overline{\g}_0}}

\newcommand{\fp}{\ensuremath{f_1}}	
\newcommand{\gp}{\ensuremath{g_1}}	
\newcommand{\Ep}{\ensuremath{E_1}}

\newcommand{\ko}{\ensuremath{k_0}}

\newcommand{\lambdaV}{\ensuremath{\lambda}} 
\newcommand{\zV}{\ensuremath{z}}	

\newcommand{\VlasovLrz}{\ensuremath{\mathcal{A}}} 
\newcommand{\VlasovLrzSD}{\ensuremath{\mathcal{B}(\theta)}}

\newcommand{\vhmin}{\ensuremath{u^{\mathrm{min}}}}
\newcommand{\vhmax}{\ensuremath{u^{\mathrm{max}}}}

\newcommand{\vador}{VADOR}

\begin{document}

\author{Evangelos Siminos} 
\author{Didier B\'enisti}
\author{Laurent Gremillet}
\affiliation{CEA, DAM, DIF, F-91297 Arpajon, France }

\title{Stability of nonlinear Vlasov-Poisson equilibria through spectral 
deformation and Fourier-Hermite expansion}

\date{\today}

\begin{abstract}
We study the stability of spatially periodic, nonlinear Vlasov-Poisson
equilibria as an eigenproblem in a Fourier-Hermite basis (in the space and
velocity variables, respectively) of finite dimension, $N$. When the advection
term in Vlasov equation is dominant, the convergence with $N$ of the eigenvalues
is rather slow, limiting the applicability of the method. We use the method of
spectral deformation introduced in [J. D. Crawford and P. D. Hislop, Ann. Phys.
{\bf 189}, 265 (1989)] to selectively damp the continuum of neutral modes
associated with the advection term, thus accelerating convergence. We validate
and benchmark the performance of our method by reproducing the kinetic
dispersion relation results for linear (spatially homogeneous) equilibria.
Finally, we study the stability of  a periodic Bernstein-Greene-Kruskal mode
with multiple phase space vortices, compare our results with numerical
simulations of the Vlasov-Poisson system and show that the initial unstable
equilibrium may evolve to different asymptotic states depending on the way it
was perturbed.
\end{abstract}


\maketitle

\section{Introduction\label{s:intro}}

The stability of stationary nonlinear electrostatic waves, 
the so-called Bernstein-Greene-Kruskal (BGK) modes 
introduced in \refref{bernstein57}, 
is a very old and basic problem that 
is still of interest\rf{paskauskas09,khain10,manfredi00}. 
It appears, moreover, to be closely related to the saturation 
of stimulated Raman scattering (SRS), 
an issue that motivated the present work. 
Indeed, for physical conditions typical of those met in 
inertial confinement fusion, the electron plasma wave (EPW) 
that grows unstable due to SRS sees its amplitude 
change over space and time scales much larger than the Debye length
and the plasma period, respectively\rf{benisti10-3}. 
In a sense, this EPW is therefore close to a BGK mode. 
Moreover, recent one-dimensional ($1$-D) Vlasov 
simulations of SRS\rf{brunner2004} indicate that Raman reflectivity stopped 
increasing monotonically with time due to the growth of sidebands, 
resulting from a purely electrostatic 
instability similar to that introduced in \refref{kruer69}. 
More recently, the so-called vortex fusion instability\rf{ghizzo1988}, 
which we will further detail in this paper, was invoked in \refref{ghizzo09} 
to explain why Raman reflectivity stopped growing monotonically.

In this paper, we strictly restrict to BGK equilibria, and describe 
a systematic and very efficient method to address their stability
properties.
Compared to a purely numerical approach consisting 
in integrating the Vlasov-Poisson system, 
our method not only very precisely 
predicts the growth rate of the
instability but also the functional form of the few fastest growing
modes. This allows one to illuminate the physics behind the instability, 
to discern different pathways for subsequent evolution depending on the
mode triggered and to devise viable control strategies\rf{sipp2010}.
Although many approaches\rf{goldman70,santini70,lewis79,ghizzo1988,paskauskas09}
have been developed to study the stability of nonlinear electrostatic waves, 
to the best of our knowledge none
has provided such a precise description of the unstable modes by using 
a very general formalism and with very moderate computational cost.

In order to determine the functional dependence of the unstable modes, 
an eigenproblem formulation is required.  
It is derived by linearizing the governing equations around \emph{any}
equilibrium distribution function $\fo(x,v)$. 
This leads to the following general formulation
\beq\label{eq:VlasovLrz}
	\frac{\partial \fp}{\partial t}=\VlasovLrz \fp\,,
\eeq
where $\VlasovLrz$ is a linear operator which depends 
on $\fo$, while $\fp(x,v,t)$ is an infinitesimal perturbation. 
The eigenvalues of \VlasovLrz\ determine the stability of the 
equilibrium characterized by $\fo$. This equilibrium is unstable if some of the 
eigenvalues of $\VlasovLrz$ have a strictly positive real part, 
the largest of which being the growth rate of the instability. 

For spatially homogeneous equilibria,
the eigenproblem  \refeq{eq:VlasovLrz} has been treated by Van
Kampen\rf{VanKampen55} and Case\rf{case59}. 
For spatially inhomogeneous BGK equilibria, characterized by
a non-vanishing electric field, we propose here an approximate resolution of 
the eigenproblem \refeq{eq:VlasovLrz} by making use of the Galerkin spectral 
method\rf{boyd2001}. Hence, we expand the total distribution function 
over a finite set of global smooth functions 
which fulfill the boundary conditions of our problem, and which are 
moreover chosen to be orthogonal. 
Then, the operator \VlasovLrz\ is approximated by a finite dimensional matrix.

In many areas of physics, most notably fluid mechanics, spectral methods have
been particularly successful in solving eigenproblems of the form 
\refeq{eq:VlasovLrz}, with exponentially fast
convergence of the result as a function of the number of orthogonal functions 
retained in the expansion for a sufficiently smooth $f$ 
(see \refref{boyd2001}, for example). 
This hallmark of spectral methods makes them 
attractive for the study of the stability of Vlasov-Poisson and Vlasov-Maxwell
equilibria, especially in dimensions higher than one.
However, a convincing application of spectral methods to the problem
of stability of BGK waves is still, to the best of our knowledge, lacking.
Problems arise owing to the very nature of the operator \VlasovLrz. In
particular the linear, advection term in Vlasov equation 
[see Eq.~\refeq{eq:Vlasov-1d} below] is responsible for the transfer of energy 
to very fine velocity scales, a phenomenon known as velocity 
space \emph{filamentation} (see, for example, \refref{manfredi97}). 
When the contribution of the advection term is significant, eigenfunctions of
\VlasovLrz\ are expected to involve high order velocity modes (fine velocity
scales) leading to slow, power law convergence of the expansions on
the orthogonal functions rather than the usual exponential convergence. 
An extreme
example is the continuum of singular, neutral van Kampen modes whose
approximation by means of smooth functions is clearly out of reach. 
Although we will not be interested in the continuum of neutral modes, it
still has to be accounted for in the numerical approximation of 
\VlasovLrz\ and can interfere with the determination of the unstable modes 
of interest, particularly the weakly unstable ones.

To ensure fast convergence of the eigenvalue calculation, it is clear that the
role of the neutral modes has to be undermined.
To this end, we use the method of \emph{spectral deformation},
originally developed for quantum mechanical problems and introduced 
in the study of the Vlasov-Poisson system by Crawford and
Hislop\rf{crawford89,hislop89}. The method introduces an operator 
\beq
    \VlasovLrzSD=\Uop{\theta}\,\VlasovLrz\,\UopI{\theta}\,,
\eeq
with $\Uop{\theta}$ non-unitary for complex $\theta$. It can be
proved\rf{hislop89} that the eigenvalues of \VlasovLrz\ 
with nonzero real part remain
unchanged under suitably chosen transformations $\Uop{\theta}$, while the
continuum of neutral modes is damped. Our central observation is that 
if the corresponding damping rate
is chosen so that the continuum spectrum is well separated from the discrete
eigenvalues of interest, then exponential convergence with the
truncation order can be recovered. 

We choose to expand the distribution function 
in Fourier series (in space) and Hermite functions (in velocity). 
This  decomposition was 
first introduced by Grant and Feix\rf{grant67} in the study of
stability of spatially homogeneous Vlasov-Poisson equilibria and
recently generalized to the full Vlasov-Maxwell system and inhomogeneous
equilibria by Camporeale \etal\rf{camporeale06}. Even more recently
Pa\v{s}kauskas and De Ninno\rf{paskauskas09} revisited the method from a nonlinear
dynamics perspective, studying homogeneous equilibria of the Hamiltonian mean
field model. We show through numerical examples that, 
with the introduction of spectral deformation,
Fourier-Hermite expansions converge fast enough to be useful for the
determination of unstable modes, even for strongly inhomogeneous equilibria.

The organization of this paper is as follows. We introduce 
the Vlasov-Poisson system and its linear approximation around an inhomogeneous
equilibrium in \refsect{s:VlasovStab}. For simplicity, we restrict 
to one space and one velocity dimension.
Spectral deformation is 
introduced and applied to the Vlasov-Poisson system in
\refsect{s:VlasovSD}. In \refsect{s:expansionFH} we derive 
the representation of \VlasovLrzSD\ in the
Fourier-Hermite basis. We illustrate the treatment of Landau damped modes
by our method in \refsect{s:landau}. In \refsect{s:bot} we compare 
our results for
a spatially homogeneous, bump-on-tail, distribution function against those obtained by
numerically solving Landau's dispersion relation. We also use
this test problem to illustrate some of the convergence issues that can arise
as the wavelength of perturbations decreases and the advection term becomes
more significant, and their resolution through spectral deformation. 
In \refsect{s:BGK} we study a nonlinear example, 
namely a BGK mode with multiple phase-space vortices. 
Our results for the growth rate and the unstable modes agree with
numerical simulations of the (nonlinear) Vlasov-Poisson system. At a qualitative
level our calculations show that the collective modes trigger the vortex fusion
observed in numerical simulations. We discuss our findings 
and the potential for optimization, as well as our 
future studies based on this work, 
in \refsect{s:conclusions}. In the appendices we
provide some technical details on the properties of the Hermite basis used here
(\refappe{s:hermite}), the representation of $U(\theta)$ in Fourier-Hermite
basis (\refappe{s:Umatrix}), the truncation of the representation of \VlasovLrz\
(\refappe{s:FHtrunc}) and the calculation of the Fourier-Hermite coefficients
(\refappe{s:FHcoef}).

\section{Linearization of the Vlasov-Poisson system\label{s:VlasovStab}}
We consider the one-dimensional motion of electrons in an unmagnetized plasma, 
with immobile ions forming a neutralizing background. 
We restrict to situations that can be modeled 
using periodic boundary conditions in a spatial domain $x\in\left[0,L\right]$. 
The motion is described in terms of the Vlasov-Poisson system
\bseq\label{eq:Vlasov-Poisson-1d}
	\beq\label{eq:Vlasov-1d}
		\frac{\partial \f}{\partial t}
		  +\vv\frac{\partial \f}{\partial x}
		      +\E\frac{\partial \f}{\partial \vv}=0\,,
	\eeq
	\beq\label{eq:Poisson-1d}
		\frac{\partial \E}{\partial x}=
		      \left(\int_{-\infty}^{+\infty}\f\, d\vv\,-1\right)\,,
	\eeq
	\beq
		\int_0^L dx\, E = 0\,,
	\eeq
\eseq
where velocity is normalized to the thermal one 
$\vv_{th}=\left(T_e/m_e\right)^{1/2}$, space to the Debye length 
$\lambda_D=\left(\epsilon_0 T_e/q_e^2 n_0\right)^{1/2}$,
time to the inverse of the electron plasma frequency 
$\omega_{pe}=\left(q_e^2 n_0/\epsilon_0 m_e\right)^{1/2}$ 
and electric field to $\vv_{th}^2 m_e/\lambda_D q_e$, 
where $q_e<0$ is the electron charge.

Let $\left[\fo(x,\vv),\Eo(x)\right]$ be an equilibrium solution of 
\refeq{eq:Vlasov-1d}, 
\beq
  \vv\frac{\partial \fo}{\partial x}+\Eo\frac{\partial \fo}{\partial \vv}=0\,.
  \label{eq:Vlasov-eqb}
\eeq
The Vlasov equation is Galilean
invariant and therefore any traveling wave solution can be reduced to an
equilibrium solution without loss of generality.

Substituting  $\f\equiv \fo+\fp\,,\ \E\equiv \Eo+\Ep$ 
in \refeq{eq:Vlasov-Poisson-1d}, where $\fp$ and $\Ep$ are infinitesimal 
perturbations, and accounting only for first order terms in $\fp$, we get
\bseq\label{eq:Vlasov-Poisson-1d-lin}
	\beq\label{eq:Vlasov-1d-lin}
		\frac{\partial \fp}{\partial t} = 
		   - \vv\frac{\partial \fp}{\partial x} 
			 - \left(\Eo\frac{\partial \fp}{\partial \vv}
				+\Ep\frac{\partial \fo}{\partial \vv}\right)\,,
	\eeq
	\beq\label{eq:Poisson-1d-lin}
		\frac{\partial \Ep}{\partial x}=
					  \int_{-\infty}^{+\infty}\,\fp\,d\vv\,,	
	\eeq
	\beq\label{eq:Div-Free-1d-lin}
		\int_0^L dx\, \Ep = 0\,,
	\eeq
\eseq
which describe the evolution of $(\fp,\Ep)$ in the linear neighborhood 
of $(\fo,\Eo)$. 

Owing to periodic boundary conditions in the space variable, the distribution 
function and electric field may be expressed in terms of Fourier series,
\bseq\label{eq:Fexp}
	\beq\label{eq:Fexpf}
		\fo(x,\vv)=\sum_{r=-\infty}^{+\infty}\fo^{r}(\vv) \Phi_r(x)\,,
	\eeq
	\beq
		\Eo(x)=\sum_{r=-\infty}^{+\infty} \Eo^r \Phi_r(x)\,,
	\eeq
\eseq
where $\Phi_r(x)\equiv e^{\ii r \ko x}$, $\ko=2\pi/L$, 
and similar expansions hold for $\fp$ and $\Ep$. Plugging system \refeq{eq:Fexp} 
into system \refeq{eq:Vlasov-Poisson-1d-lin}
and using the standard Fourier basis orthogonality relations, we get
\bseq\label{eq:Vlasov-Poisson-1d-lin-F}
\begin{multline}\label{eq:Vlasov-1d-lin-F}
      \frac{\partial}{\partial t} \fp^{k}(\vv,t) = 
		      -\ii k \ko \vv \fp^{k}(\vv,t)\brk
		+\frac{\ii}{k_{0}}\sump_{r=-\infty}^{\infty} 
		  \frac{1}{r}\frac{\partial}{\partial \vv}\fo^{k-r}(\vv)
		    \int_{-\infty}^{+\infty}\,d\vv\,\fp^{r}(\vv,t)\brk
		+\frac{\ii}{k_{0}}\sump_{r=-\infty}^{\infty}\frac{1}{r}
		  \frac{\partial}{\partial \vv}\fp^{k-r}(\vv,t)
			\int_{-\infty}^{+\infty}\,d\vv\,\fo^{r}(\vv)\,,
\end{multline}
\beq\label{eq:Poisson-1d-lin-F}
  \Ep^{k}(t) =	
	\begin{cases}
	  0\,, 	& \text{if $k=0$,}\\
	  -\frac{\ii}{k \ko}\int_{-\infty}^{+\infty}\,d\vv\, \fp^{k}(\vv,t)\,, 
		& \text{if $k\neq 0$,}
	\end{cases}
\eeq
\eseq
where the prime in summations indicates that we omit the $r=0$ term, as we have 
incorporated \refeq{eq:Poisson-1d-lin-F} into \refeq{eq:Vlasov-1d-lin-F} 
to eliminate the electric field.
The restriction $\Ep^{0}=0$ follows from condition 
\refeq{eq:Div-Free-1d-lin} on the electric field.
Equation 
\refeq{eq:Vlasov-1d-lin-F} is of the form \refeq{eq:VlasovLrz},
\[
 \frac{\partial \fp}{\partial t}=\VlasovLrz \fp\,,
\]
where \VlasovLrz\ is a linear integro-differential operator 
that depends on $\fo$.
For the rest of this paper we will use the distribution function alone to refer 
to solutions, keeping
in mind that the electric field is determined self-consistently through 
Poisson's equation.

\begin{figure}
 (a)~\includegraphics[width=0.21\textwidth,clip=true]{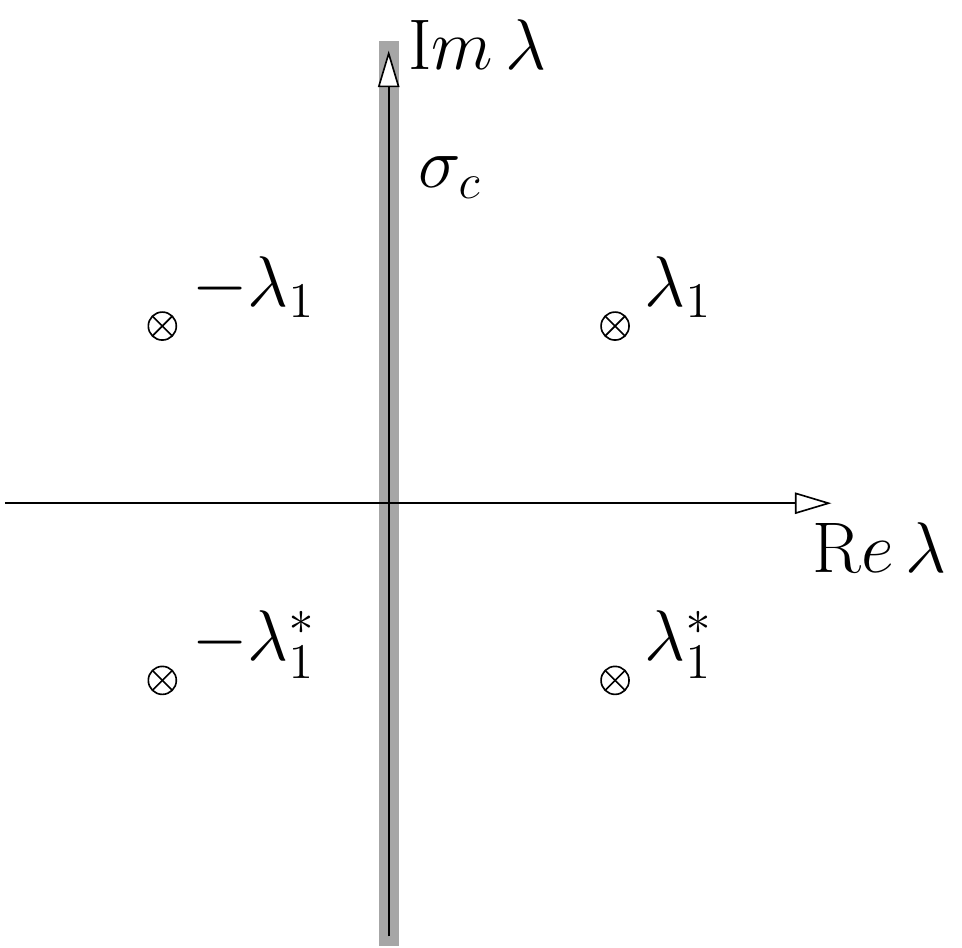}
 (b)~\includegraphics[width=0.21\textwidth,clip=true]{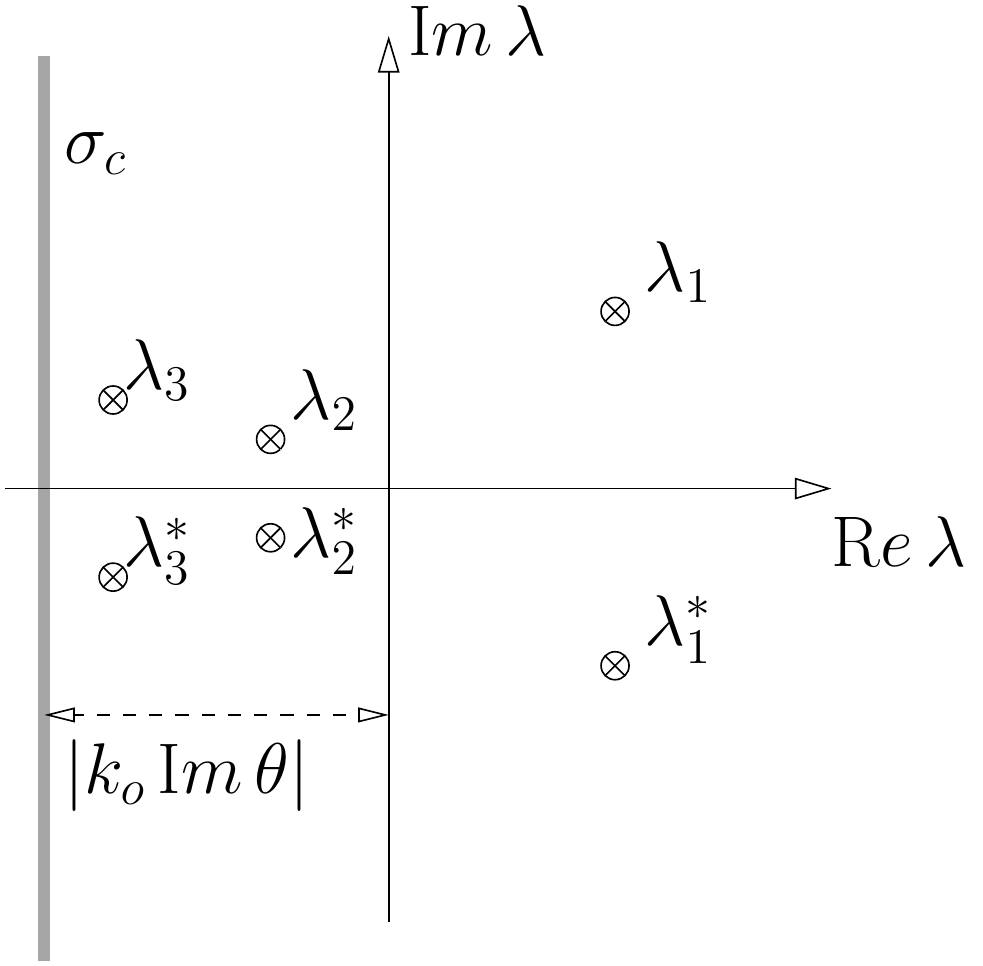}
\caption{
  (a) Spectrum of an unstable, spatially homogeneous Vlasov equilibrium: 
	$\pm\lambdaV_1$ and
  $\pm\lambdaV_1^*$ are discrete eigenvalues, while $\sigma_{\mathrm{c}}$ 
  is the continuous spectrum (gray, thickened for clarity). 
  (b) Through spectral deformation, 
  $\sigma_{\mathrm{c}}$
  is shifted to the left half-plane, uncovering eigenvalues 
  $\lambdaV_2,\lambdaV_2^*$ and $\lambdaV_3,\lambdaV_3^*$ 
  corresponding to damped modes  
  [$-\Re(\lambdaV_i)$ being the damping rate obtained through
	 Landau's dielectric tensor formalism].
  Since the Hamiltonian structure is destroyed by
  spectral deformation, $-\lambdaV_1,-\lambdaV_1^*$ disappear as 
  $\sigma_{\mathrm{c}}$ moves to their left (after \refref{crawford89}).}
\label{f:SD}
\end{figure}

Owing to the Hamiltonian structure of the Vlasov-Poisson system\rf{morrison80}, 
eigenvalues $\lambdaV_n\equiv \gamma_n+\ii \omega_n$ 
of the real operator \VlasovLrz\ come into quartets 
$\pm\lambdaV_i,\,\pm\lambdaV_i^*$ [see \reffig{f:SD}(a)].
Moreover, $\VlasovLrz$ characteristically has a continuum spectrum $\sigma_c$
on the imaginary axis.

If $\gamma_n\equiv \Re(\lambdaV_n)>0$ for some $\lambdaV_n\in\sigma$, 
a perturbation in the $n$'th eigenspace grows in modulus   
as $e^{\gamma_n t}$, 
while its phase oscillates at frequency $\omega_n\equiv\Im (\lambda_n)$. 
Then, the norm of a generic perturbation 
having non-zero components along all eigendirections 
would grow asymptotically in time 
as $e^{\max(\gamma_n)t}$. In this case, we will say that 
the equilibrium is unstable.

If no eigenvalue with strictly positive real part exists,  
$\Re(\lambdaV_n)=0$ for all $n$, the eigenvalues form
a continuoum which coincides with the imaginary axis and
the corresponding eigenmodes are undamped (neutrally stable). 
These modes are singular (described by generalized functions or distributions) 
and do not represent physically observable modes of the system. However, their
presence is connected to the collisionless damping of generic electric 
field perturbations. 
Solving the initial value problem for small amplitude electrostatic waves 
in a Maxwellian plasma, Landau\rf{landau46} famously showed that
the electric field amplitude vanishes at an exponential rate. 
As shown by Van Kampen\rf{VanKampen55} and generalized to more general spatially
homogeneous equilibria by Case\rf{case59}, Landau damping can be understood as a
destructive interference effect (known as \emph{phase mixing}) of the
neutral modes. 

In this paper, we will adopt the convention of sorting our eigenvalues 
by decreasing real part, so
that for unstable equilibria $\gamma_1$ corresponds to the largest growth rate
and the eigenfunction $\ev{1}$ is the fastest growing mode. Moreover, 
we will not use  $\zV=\ii\lambdaV_i$ as is traditionally the case 
in plasma physics when
writing solutions of \refeq{eq:Vlasov-Poisson-1d-lin}. 
Instead, we conform to the convention usually employed 
in the study of linear ordinary differential equations, 
which is more natural to a spectral
discretization of the eigenproblem for $\VlasovLrz$. 
As a result, the continuous part of the spectrum, $\sigma_{\mathrm{c}}$, 
lies on the imaginary axis, as illustrated 
in \reffig{f:SD}(a). 

\section{Spectral deformation\label{s:VlasovSD}}

As already stated in the introduction, in order to alleviate the difficulty 
due to the continuum of neutral modes in the spectrum of \VlasovLrz, 
we do not address the stability of a BGK equilibrium by solving the eigenproblem 
for \VlasovLrz, but for the transformed operator
\beq\label{eq:SD}
    \VlasovLrzSD=\Uop{\theta}\,\VlasovLrz\,\UopI{\theta}, 
\eeq
with
\beq\label{eq:VelTrans}
	\Uop{\theta}h^{k}(v)=h^k(v+\theta_k)\,.
\eeq
Here,
\[
\theta_k=\begin{cases}
          	\sgn(k)\theta, & k\neq0\,,\\
		0, 	& k=0\,,
         \end{cases}
\]
and $h^k(v)$ is the $k$th Fourier mode of a given function $h(x,v)$. 
As shown by Hislop and Crawford in Refs.~\cite{crawford89,hislop89}, 
the main merit of this transformation is that for $\Im(\theta)<0$ the continuous
spectrum becomes damped, while the eigenvalues with  
$\Re(\lambda)>0$ remain unchanged 
[see \reffig{f:SD}(b) and \refref{crawford89} for a homogenous equilibrium].
When eigenvalues with $\Re(\lambda)=0$ of multiplicity higher than one
exist, we can talk of discrete eigenvalues
``embedded'' in the continuous spectrum (see \refref{crawford89}). These
embedded discrete eigenvalues also remain unchanged under spectral deformation.
Hence, the stability issue of a BGK equilibrium 
may be equivalently  addressed by calculating the eigenvalues of \VlasovLrz~or 
of \VlasovLrzSD, except that these are much more easily and accurately 
estimated for \VlasovLrzSD, since the damped continuous spectrum is 
much easier to account for numerically.

One may however wonder how Landau damping, which results from the phase mixing 
of the neutral modes, could be recovered using spectral deformation.~As shown
in \refref{crawford89}, for a homogeneous equilibrium, when $\Im(\theta)<0$ 
the neutral modes of $\mathcal{A}$ become damped by 
$\ko \Im(\theta)$. When this is larger in absolute value than the 
damping rate $\gamma_n$ obtained through Landau's dielectric tensor formalism, 
a new pair of discrete eigenvalues $\lambda_n,\, \lambda_n^*$ 
with $|\Re(\lambda_n)|=\gamma_n$ [and $\Im(\lambda_n)$ equal to
the frequency predicted by Landau] appears in the spectrum 
of \VlasovLrzSD~[see \reffig{f:SD}(b)]. The complex conjugate eigenvalue
corresponds to wavelength $-\ko$. Hence, Landau damping is indeed 
recovered but, unlike for the original Vlasov-Poisson system, it now appears 
as being due to the damping of an eigenmode (corresponding to the largest
negative eigenvalue) for the dissipative dynamics 
represented by the operator \VlasovLrzSD. This will be discussed in more 
detail in \refsect{s:landau}.

Let us now, as for the original Vlasov-Poisson system, 
write the eigenproblem for \VlasovLrzSD~in Fourier space. 
Using, \refeq{eq:Vlasov-1d-lin-F} and $g(v)\equiv\Uop{\theta}f(v)$, 
one easily finds that the equation 
\beq\label{eq:VlasovLrzSD}
	 \frac{\partial \gp}{\partial t}=\VlasovLrzSD \gp
\eeq
becomes, in Fourier space,
\small
\begin{multline}\label{eq:Vlasov-1d-lin-F-g} 
      \frac{\partial}{\partial t} \gp^{k}(\vv,t) = 
	    -\ii k \ko (\vv+\theta_k) \gp^{k}(\vv,t)\\
		+\frac{\ii}{\ko}\sump_{r=-\infty}^{\infty} 
		    \frac{1}{r}\frac{\partial}{\partial \vv}
		      \go^{k-r}(\vv+\theta_k-\theta_{k-r})
			  \int_{-\infty}^{+\infty}\,d\vv\,\gp^{r}(\vv,t)\brk
		+\frac{\ii}{\ko}\sump_{r=-\infty}^{\infty}\frac{1}{r}
		 \Uop{\theta_k-\theta_{k-r}}\frac{\partial}{\partial \vv}
		  \gp^{k-r}(\vv,t)\int_{-\infty}^{+\infty}\,d\vv\,\go^{r}(\vv).
\end{multline}
\normalsize
In \refeq{eq:Vlasov-1d-lin-F-g} 
the introduction of dissipation through spectral 
deformation with $\Im(\theta)<0$ becomes apparent 
through the presence of $\theta_k$ in the advection term. 
The invariance of the discrete spectrum with 
$\Re(\lambda)\geq0$ is not obvious from \refeq{eq:Vlasov-1d-lin-F-g} 
but has been established in
\refref{hislop89}. The unstable eigenmodes of the initial Vlasov-Poisson 
problem can be recovered through the inverse transformation, 
$\fp^k(v,t)=\Uop{-\theta_k}\gp^k(v,t)$, 
since $\partial_t (U^{-1}g)=\mathcal{A}(U^{-1}g)$ and $\Uop{\theta}$ is
time independent.

\section{Hermite expansion\label{s:expansionFH}}

In order to approximately solve the eigenproblem for \VlasovLrzSD, 
we now write $g^k(v)$ as a sum of orthonormal Hermite functions. 
Such expansions have been first introduced in numerical studies of the
Vlasov-Poisson system by Grant and Feix\rf{grant67}
and Armstrong\rf{armstrong67} and present various advantages.  
Hermite functions decay as $e^{-\vv^2}$ at large $\vv$ and therefore
allow one to treat boundary conditions correctly without truncating
the infinite interval. Moreover, convenient three term relations 
(see \refappe{s:hermite}) 
of the Hermite functions will result in an explicit, sparse 
representation of the operator $\VlasovLrzSD$.
Hermite functions are related to velocity 
moments, the few first of which are directly linked to physical 
observables\rf{grant67,camporeale06}.
As pointed out by Pa\v{s}kauskas and De Ninno\rf{paskauskas09},
this allows one, at least in principle, to naturally separate 
thermal and filamentation scale effects. 
As we will see in our numerical examples of \refsect{s:stabResults}, 
filamentation scale 
often strikes back, rendering Hermite expansions problematic, a shortcoming
pointed out a long time ago\rf{grant67} and
overcome here by the introduction of spectral deformation.

Here we consider the so-called \emph{asymmetrically weighted}
Hermite basis\rf{schumer98}. 
Denoting by $\Psi_n(v)$ the basis functions and by $\Psi^n(v)$ the weight 
functions we have
\beq\label{eq:HermiteFunc}
	\Psi_n(v)=C_n e^{-v^2}H_n(v)\,,\quad \Psi^n(v)= C_n H_n(v)\,,
\eeq
where $H_n(v)$ are Hermite polynomials and $C_n=1/(\pi^{1/4}\sqrt{2^n n!})$. 
More details on the properties of Hermite polynomials
used here can be found in \refappe{s:hermite}. 
We note the 
important orthonormality relation
\beq
	\int_{-\infty}^{+\infty} \Psi^m(v)\Psi_n(v)\,dv=\delta_{mn}\,.
\eeq

Our expansion of $g^k(v,t)$ over the Hermite functions reads
\beq\label{eq:Hexpg}
 	g^k(\vv,t)=\sum_{s=0}^{+\infty}g^{ks}(t)\Psi_s(\vh)\,,
\eeq
where $\vv\equiv \vs \vh$, with $\vs$ an arbitrary velocity scale factor
whose importance will be discussed at the end of this section.
Plugging Eq.~(\ref{eq:Hexpg}) into \refeq{eq:Vlasov-1d-lin-F-g}, 
multiplying by $\Psi^{j}(\vh)$, integrating over $\vh$ and using 
\refeq{eq:HermiteRec}, 
\refeq{eq:innerPsiShift} and the orthonormality of the Hermite basis we get, 
when $j\geq1$,
\begin{widetext}
	\begin{align}\label{eq:Vlasov-1d-lin-FH-g}
      \frac{d}{dt} \gp^{kj} &= 
	      -\ii k \ko\left[\vs\left(\sqrt{\frac{j}{2}}\gp^{k,j-1}+\sqrt{\frac{j+1}{2}}\gp^{k,j+1}\right)
			  +\theta_k \gp^{kj}\right]\\
		&\quad -\frac{\ii \pi^{1/4}}{\ko}\sump_{r=-\infty}^{\infty}\frac{\sqrt{2j}}{r}\goTransl^{k-r,j-1}
				      \gp^{r0}
			  -\frac{\ii \pi^{1/4}}{\ko}\sump_{r=-\infty}^{\infty}\frac{\go^{r0}}{r}
				      \sum_{n=0}^{j-1}K_{j,n+1}\left(\theta_k-\theta_{k-r}\right)\gp^{k-r,n}\,,
	\end{align}
\end{widetext}
while
\beq
\frac{d g_1^{k0}}{dt} =-\ii k\ko\left[\frac{v_s}{\sqrt{2}} g_1^{k1}+\theta_k g_1^{k0} \right].
\eeq

In Eq. (\ref{eq:Vlasov-1d-lin-FH-g}) we have introduced the notations
\footnotesize
\beq
K_{jn}\left(y\right)=	\begin{cases}
					(-2)^{j-n}\sqrt{2n}\frac{C_j}{C_{n}}\binom{j}{n}\left(\frac{y}{\vs}\right)^{j-n}\,, &
												\text{if $j<n$}, \\
					\sqrt{2n}\,, & \text{if $j=n$}, \\ 
					0\,, & \text{if $j>n$,} 
				\end{cases}
\eeq
\normalsize
and $\goTransl^{k-r}(v)\equiv \go^{k-r}(v+\theta_k-\theta_{k-r})=\fo^{k-r}(v+\theta_{k})$.

We can write \refeq{eq:Vlasov-1d-lin-FH-g} in tensorial form as
\beq
	\frac{d}{dt}\gp^{kj}= \sum_{l,m} B^{kj}{}_{lm}\, \gp^{lm}\,,
\eeq
where, 
\beq
\label{did2}
	B^{k0}{}_{lm}=-ik k_0 \delta_{kl}\left[\frac{v_s}{\sqrt{2}}\delta_{1m}+\theta_k \delta_{0m}\right],
\eeq
while, when $j\geq1$, 
\beq\label{eq:VlasovLrz-SD-FH}
B^{kj}{}_{lm} \equiv D^{kj}{}_{lm}+ F^{kj}{}_{lm}
				+G^{kj}{}_{lm}\,, 
\eeq
and we have introduced
\begin{widetext}
\begin{align}
  D^{kj}{}_{lm} &= 
		-\ii k \ko \delta_{kl}\left[\vs 
		    \left(\sqrt{\frac{j}{2}}\delta_{j,m+1}
				+\sqrt{\frac{j+1}{2}}\delta_{j,m-1}\right)
					  +\theta_k\delta_{jm}\right]\,,\label{eq:D}\\ 
  F^{kj}{}_{lm} &= \begin{cases}
	-\ii\frac{\pi^{1/4}}{\ko} 
	   \frac{\sqrt{2j}}{l} \delta_{0m} \goTransl^{k-l,j-1}\,, & \text{if $l\neq0$}\,,\\
	0\,, & \text{if $l=0$\,,}
                \end{cases}\label{eq:F}\\
  G^{kj}{}_{lm} &= \begin{cases}
	- \ii\frac{\pi^{1/4}}{\ko} 
	    \frac{\go^{k-l,0}}{k-l} K_{j,m+1}\left(\theta_{k}-\theta_{l}\right),
	 		& \text{if $l\neq k$ and $m\leq j-1$}\,,\\
	0\,, 		& \text{otherwise.}
                \end{cases}\label{eq:G}
\end{align}
\end{widetext}

Equations \refeqs{did2}{eq:G} provide the representation 
of the linear operator $\VlasovLrzSD$ in the Fourier-Hermite basis.
In practice, we find it more convenient and efficient to compute 
the Fourier-Hermite coefficients $\fo^{kj}$ of $\fo(x,v)$ rather than
that of $\go(x,v)$. Then, $\go^{k-l,0}=\fo^{k-l,0}$ and we can compute
$\goTransl^{k-r,j-1}=\left(\Uop{\theta_{k}}\fo\right)^{k-r,j-1}$
as described in \refappe{s:Umatrix}.

\begin{figure}
  (a)~\includegraphics[width=0.2\textwidth]{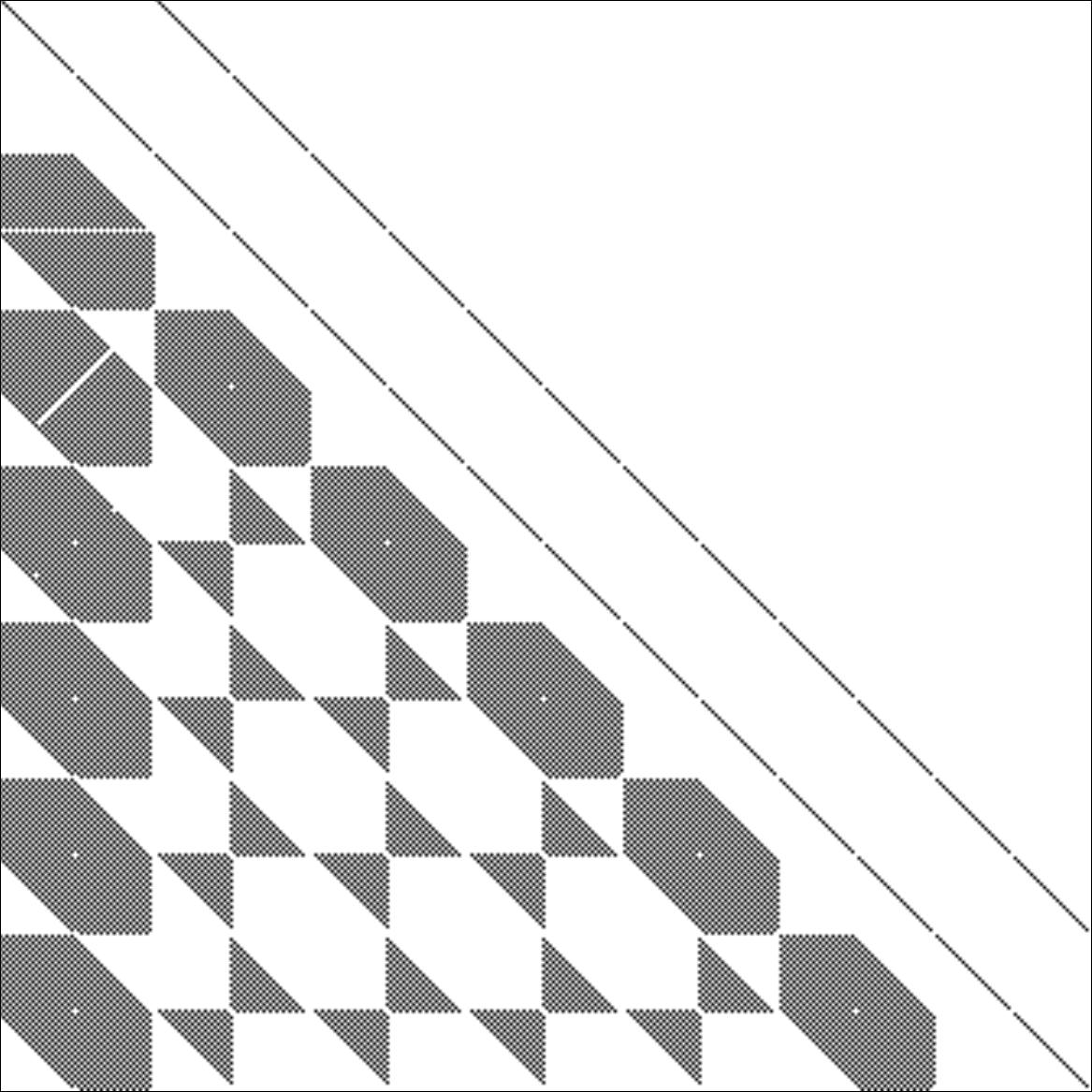}~
  (b)~\includegraphics[width=0.2\textwidth]{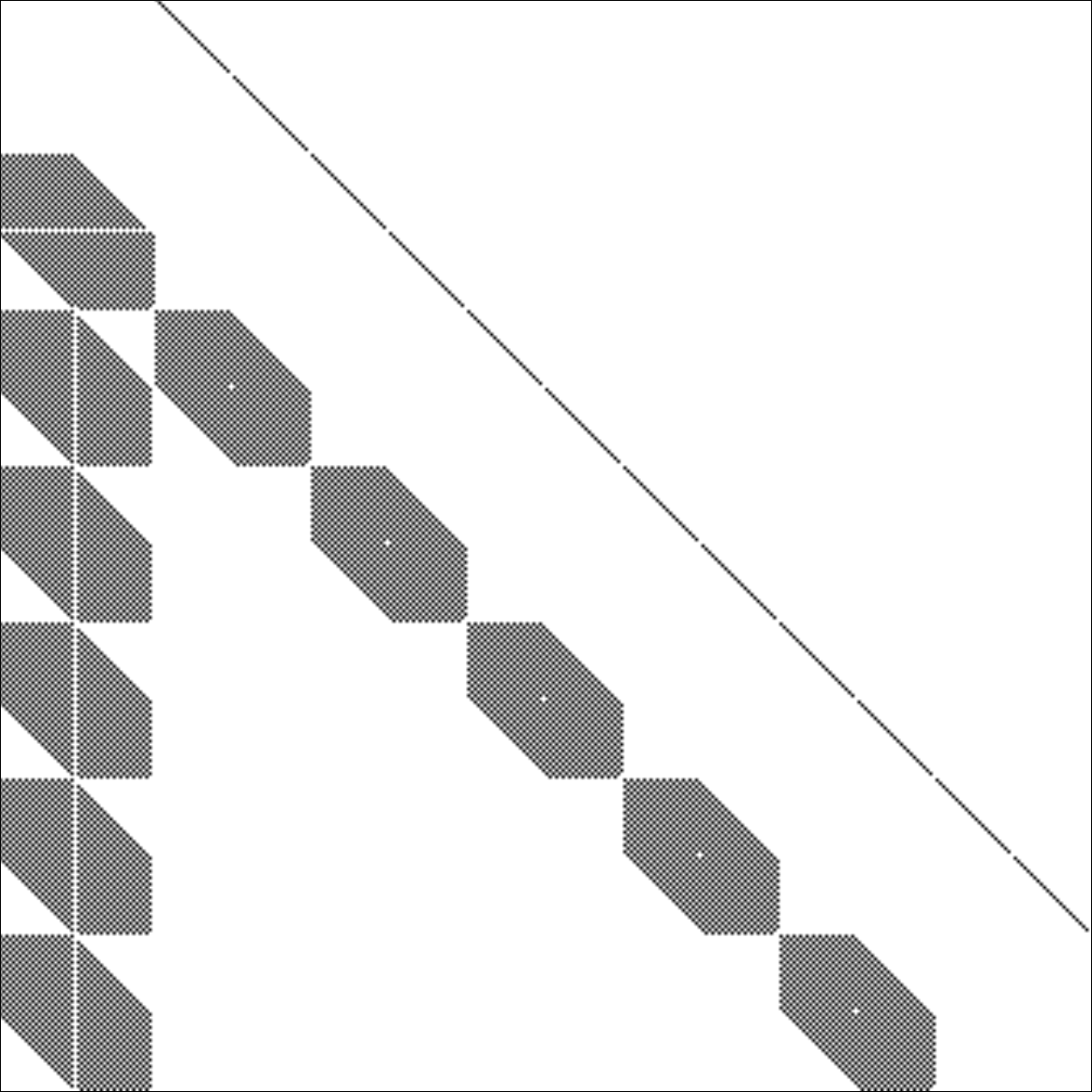}
  \caption{
		Structure of $B^{kj}{}_{lm}$, for 
		(a) $\theta\neq0$ and 
	    	(b) $\theta=0$, see \refeq{eq:VlasovLrz-SD-FH} 
		and \refeq{eq:VlasovLrz-FH}, respectively. 
		We store $B^{kj}{}_{lm}$ in a matrix 
		as described in \refappe{s:FHtrunc}. 
		White space denotes zero elements.}
  \label{f:sparse}
\end{figure}

In computations, $B^{kj}{}_{lm}$ has to be truncated to finite order
by setting $f^{N_x+1,j}(t)=f^{k,N_v+1}(t)=0$ for some
cutoff values $N_x$ and $N_v$, see \refappe{s:FHtrunc}.
Physically, such a truncation holds provided that $\fo$ \emph{and} $\fp$
can be well described by functions that do not oscillate too rapidly
with space and velocity. 
The truncated matrix, shown in \reffig{f:sparse}(a), is sparse
and the computation of its first few eigenvalues with 
the largest real parts can be efficiently handled by
iterative schemes, such as Arnoldi iteration\rf{trefethen97}. 
For $\theta=0$, $A^{kj}{}_{lm}$ contains even fewer
non-zero elements, see \refappe{s:FHtrunc} and \reffig{f:sparse}(b).

The velocity factor $\vs$ reflects the freedom to rescale the infinite domain
in which Hermite functions are defined. This freedom has been exploited 
both in numerical simulations\rf{schumer98} and eigenproblem 
calculations\rf{camporeale06, paskauskas09}. Pa\v{s}kauskas and De Ninno
use $\vs$ to minimize the number of modes needed for a good
approximation of the unperturbed distribution function $\fo$. 
Camporeale \etal,
on the other hand, optimize $\vs$ 
(and also allow for a shift of origin in $\vh$) 
iteratively so as to reduce the quadratic error
in the eigenvectors $\fp$, given an initial approximation 
computed with non-optimal $\vs$.
We have found that for the problems of interest
to us, \ie\ strongly inhomogeneous equilibria,
varying $v_s$ can be used to ensure fast convergence of the expansion
of $\fo$. However, in cases of power-law convergence of the eigenvalue
computation, we have found that tuning $v_s$ is of limited help as it
would not yield exponentialy fast convergence.
Therefore, in our numerical examples we will 
fix $\vs$ to a value that provides fast convergence of the expansion of $\fo$ 
and resort to spectral deformation to ensure exponential convergence rate 
of the eigenvalue calculation.

\section{Numerical Results\label{s:stabResults}}
\subsection{Comparison with dielelectric tensor formalism}

For spatially homogeneous unperturbed distribution functions, the equilibrium
electric field vanishes. The problem decouples in Fourier space and the growth
(or damping) rate and frequency vary continuously with $k$ and can be computed
by Landau's dielelectric tensor formalism\rf{landau46}. We will test the
validity of our expansions by comparing our results against the predictions of
Landau's theory for two examples. In the first example, $f_0$ is a Maxwellian,
which lets us discuss  the qualitative difference in addressing Landau damping
when diagonalizing $\VlasovLrz$ versus $\VlasovLrzSD$. In the second example, we
extensively compare the rate of convergence and the accuracy obtained with and
without spectral deformation for an unstable ``bump-on-tail'' distribution.

\subsubsection{Landau Damping\label{s:landau}}

\begin{figure*}
  \begin{center}
  	(a)\includegraphics[width=0.3\textwidth]{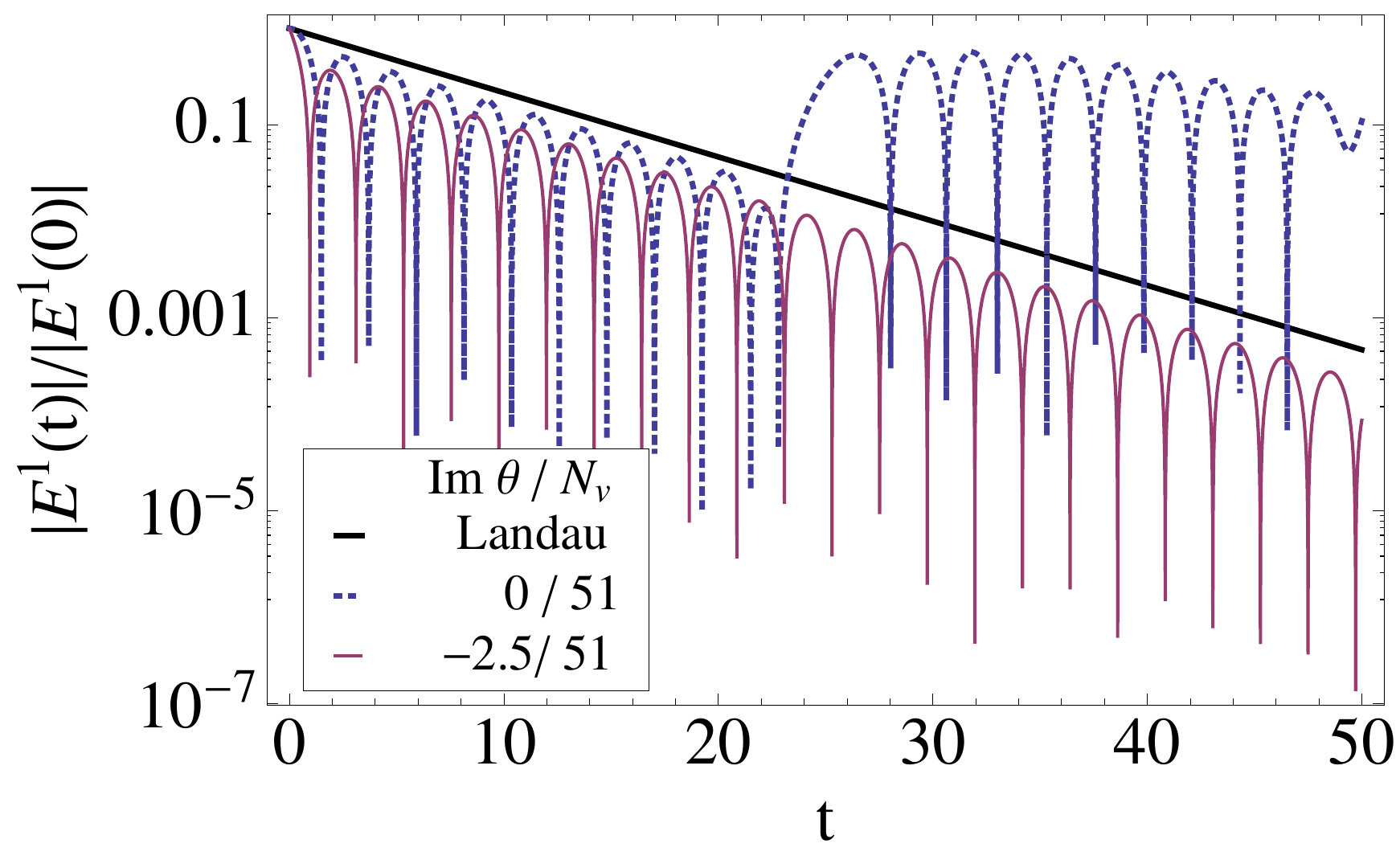}~
  	(b)\includegraphics[width=0.3\textwidth]{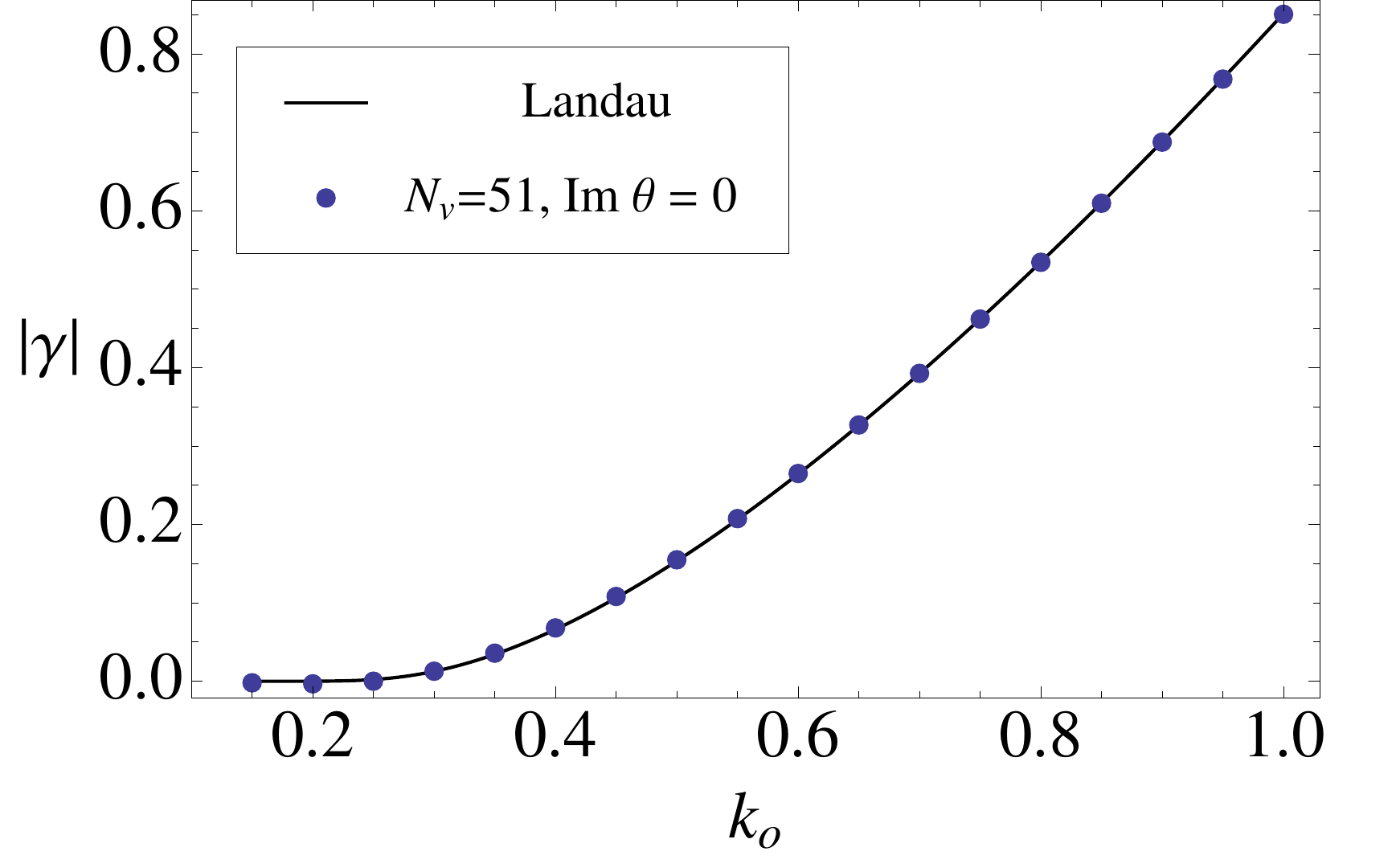}~
	(c)\includegraphics[width=0.3\textwidth]{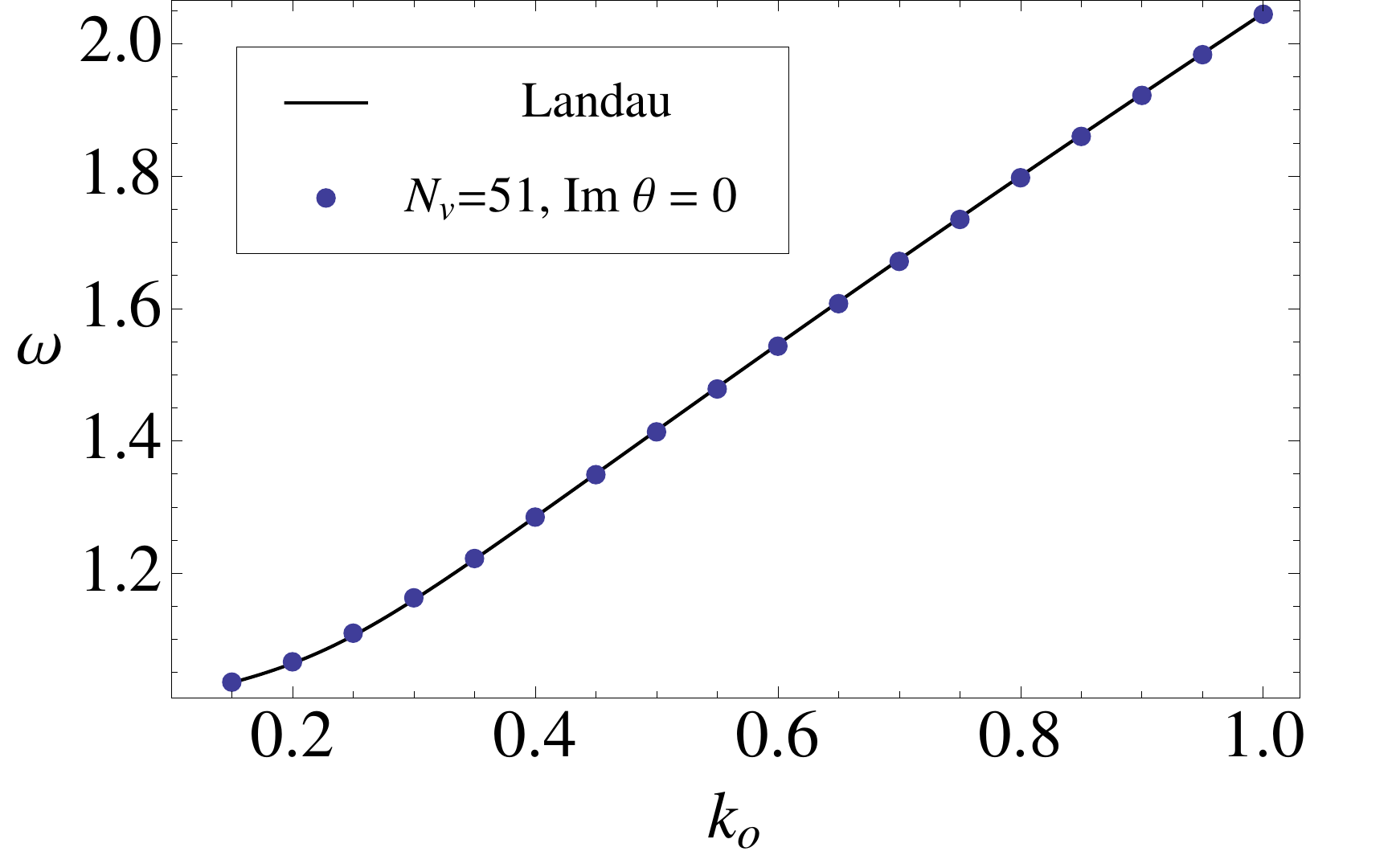}
  \end{center}
  \caption{(Color online)
	(a) Landau damping of an EPW as prescribed by Eq.~\refeq{eq:landau-IVP}, 
	with $N_v=51,\,\vs=\sqrt{2}$ (blue dotted line), 
	spectral deformation with 
	$\theta=-2.5\ii,\, N_v=51,\,\vs=\sqrt{2}$ (red solid line) and Landau's 
	prediction for the damping of the envelope (black solid line).
  	The recurrence effect is clearly seen in the blue dotted line
	for $\Im(\theta)=0$. 
	(b) Landau damping rate and (c) frequency of oscillation of
	an EPW as predicted by Landau's analysis (black solid line) and 
	Fourier-Hermite expansion with 
	$N_v=51,\, \vs=\sqrt{2},\, \Im(\theta)=0$ (blue dots).
	}
\label{f:landau}
\end{figure*}
Since we approximate the linear operator \VlasovLrz\ by a finite dimensional
matrix, the continuous, neutral spectrum is represented by a discrete set of
eigenvalues, the separation of which decreases as we increase $N_v$.
As shown by Grant and Feix\rf{grant67}, Landau damping is then only 
approximately recovered when solving  the initial value problem for the electric
field. Indeed, if we consider a Maxwellian $\fo(\vv)$ and 
an initial small amplitude electric field 
perturbation of the form $\Ep(x,0)\sim\epsilon \cos \ko x$, which is not an eigenmode 
of the problem, 
then the solution of the initial value problem in the linear approximation 
yields for the first Fourier component, $\Ep^1$, of the electric field\rf{grant67}
\beq\label{eq:landau-IVP}
	\Ep^1(t)=\Ep^1(0)\sum_{j=0}^{N_v}(a^{-1})_{j0}a_{0j} e^{\ii\omega_j t}\,,
\eeq 
where the 
$\omega_j$'s are the eigenvalues of $A^{kj}{}_{lm}$ [given by 
\refeq{eq:VlasovLrz-FH}], 
 $a_{ij}$ the matrix of its column eigenvectors, and
$(a^{-1})_{ij}$ the matrix of row vectors of the dual basis.
Hence, Landau damping of the EPW occurs as the consequence of 
the destructive interference effect 
of the neutral eigenmodes, but only over 
a finite time, because the sum (\ref{eq:landau-IVP}) is finite. 
This is illustrated in \refFig{f:landau}(a) plotting  the time evolution 
of the EPW amplitude as predicted by \refeq{eq:landau-IVP},
for $\ko=0.5$. The electrostatic wave is indeed initially damped and the damping 
rate and frequency of oscillation can be determined
from the slope and local maxima distance in \reffig{f:landau}(a),
$\gamma\simeq-0.1534$, $\omega\simeq1.416$. 
As shown in \reffig{f:landau}(b) and (c), agreement with the results
from Landau's analysis is excellent. 
However, after $t\simeq24$ 
the electric field amplitude grows again
to finite magnitude, a fact known as  \emph{recurrence}. 
The \emph{recurrence time} 
increases as we decrease the scale factor $\vs$ or we increase $N_v$
(see \refref{schumer98}) indicating that recurrence is ultimately related to the
approximation of the continuous spectrum $\sigma_c$ by a finite set
of eigenvalues (see also \refref{Knorr74}).

With spectral deformation, $\theta=-2.5\ii$, the physics of the eigenproblem
becomes dissipative: the Landau damped mode appears in the spectrum as a true 
eigenmode with $\lambda_1\simeq  -0.1534 + 1.416\,\ii$, 
accompanied by $\lambda_1^*$ and recurrence is absent. 
At the same time some of the collisionless
physics is sacrificed, namely phenomena that depend on the reversible
structure of Vlasov equation and the creation of fine velocity
scales in the distribution function, such as plasma 
echos~(See \refref{chen95}, for example). We do not
plot the results obtained with spectral deformation in \reffig{f:landau}(b) and
(c) since, within the resolution of this plot, they perfectly overlap with the
$\Im(\theta)=0$ results. 
 

\subsubsection{\label{s:bot} Bump-on-tail instability}

\begin{figure*}
  \begin{center}
  	(a)\includegraphics[width=0.3\textwidth,clip=true]{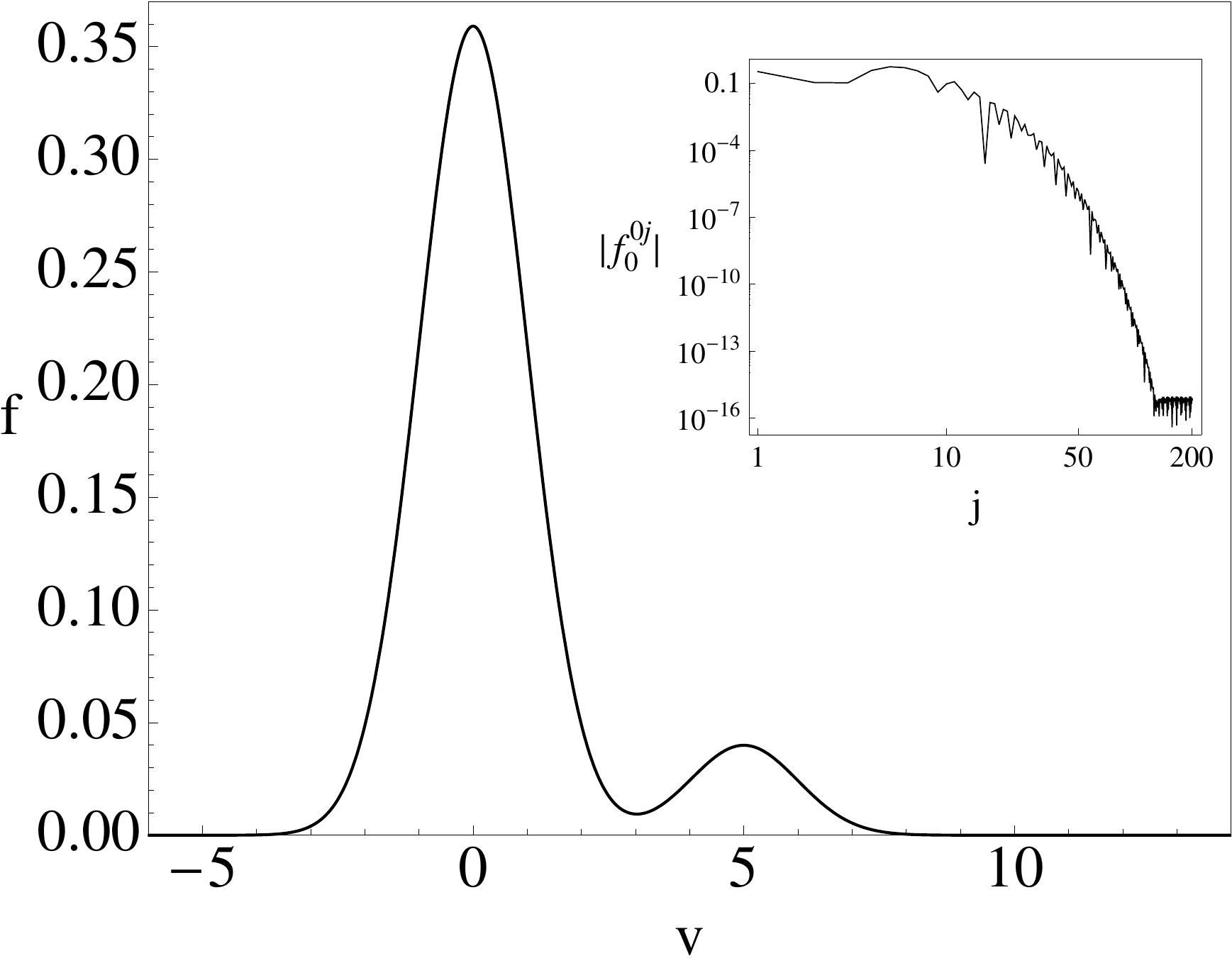}~
  	(b)\includegraphics[width=0.3\textwidth,clip=true]{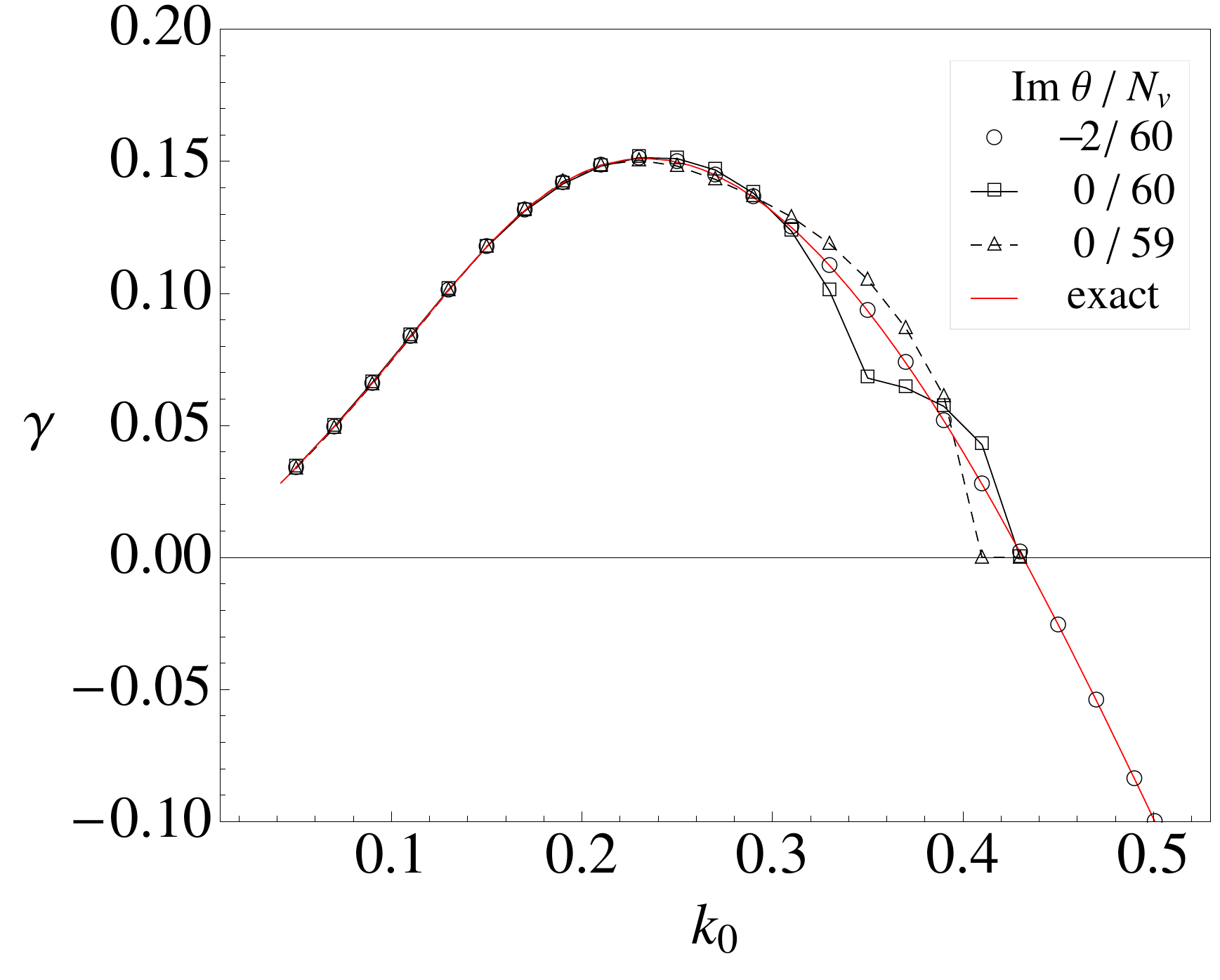}~
   	(c)\includegraphics[width=0.3\textwidth,clip=true]{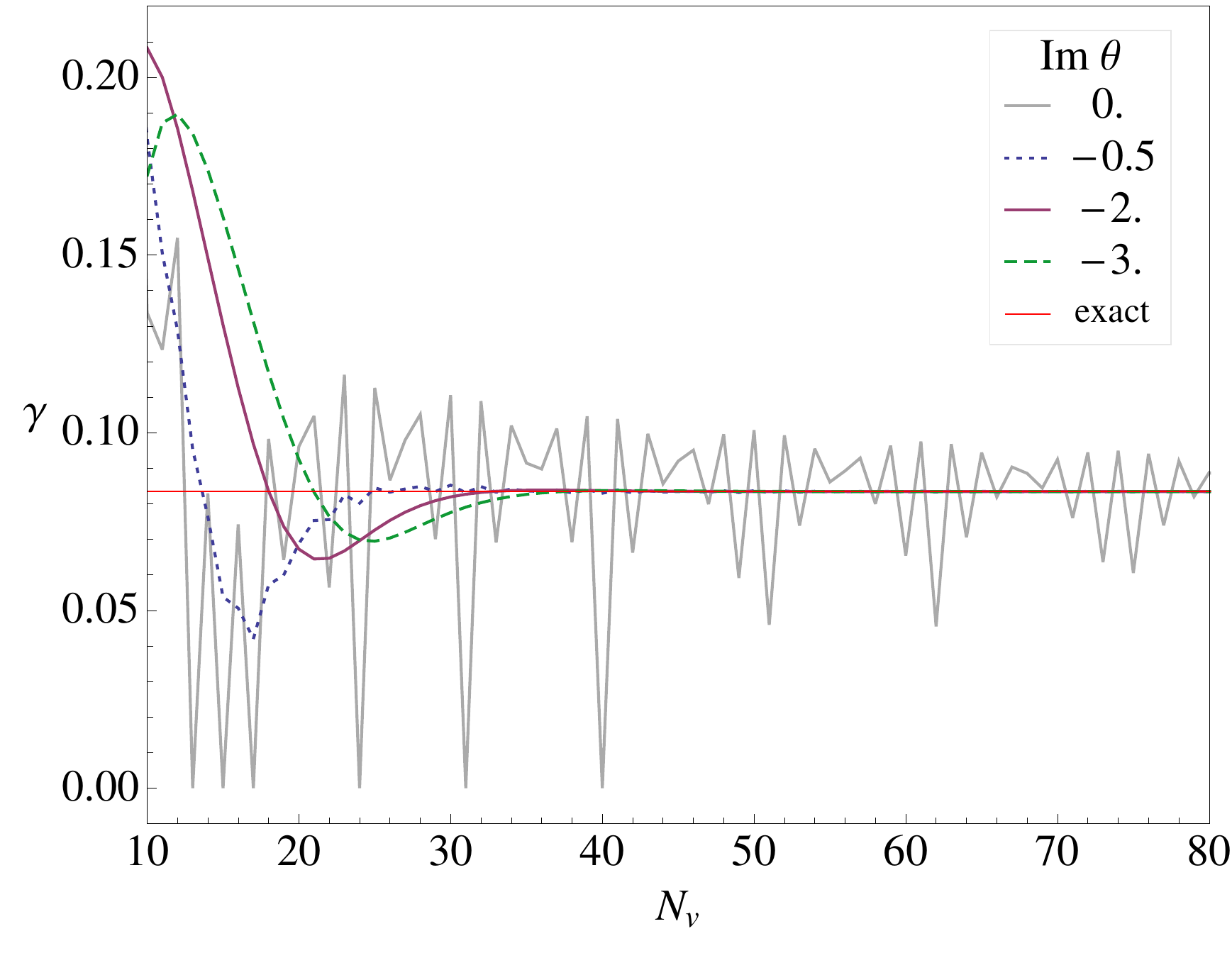}
  \end{center}
  \caption{(Color online) 
    (a) Bump-on-tail distribution \refeq{eq:bot} with 
	$n_p=0.9,\, n_b=0.1,\, v_{b}=5$ and the magnitude of
     	its Hermite expansion coefficients $|\fo^{0j}|$ in the inset.
    (b) Growth rate $\gamma$ as a function of $\ko$ as predicted by
	Landau's method (red solid line) and by Hermite expansion, with and
	without spectral deformation (black lines).
    (c) Variation of the growth rate $\gamma$ with increasing $N_v$, compared
	to the exact value for $\ko=0.36$ and $\theta=0,-0.5\ii,-2\ii,-3\ii$.
		}
\label{f:bot}
\end{figure*}

Let us now compare the results of our method against those obtained 
from the numerical 
resolution of Landau's dielectric tensor formalism 
for a bump-on-tail distribution of the form
\beq\label{eq:bot}
 \f(\vv)=\frac{n_p}{\sqrt{2\pi}}e^{-\vv^2/2}
   +\frac{n_b}{\sqrt{2\pi}}e^{-(\vv-\vv_{b})^2/2}\,,
\eeq
with $n_p = 0.9,\, n_b = 0.1$ and $\vv_{b} = 5$, see \reffig{f:bot}(a).

\begin{figure*}
  \begin{center}
   (a)\includegraphics[width=0.3\textwidth,clip=true]{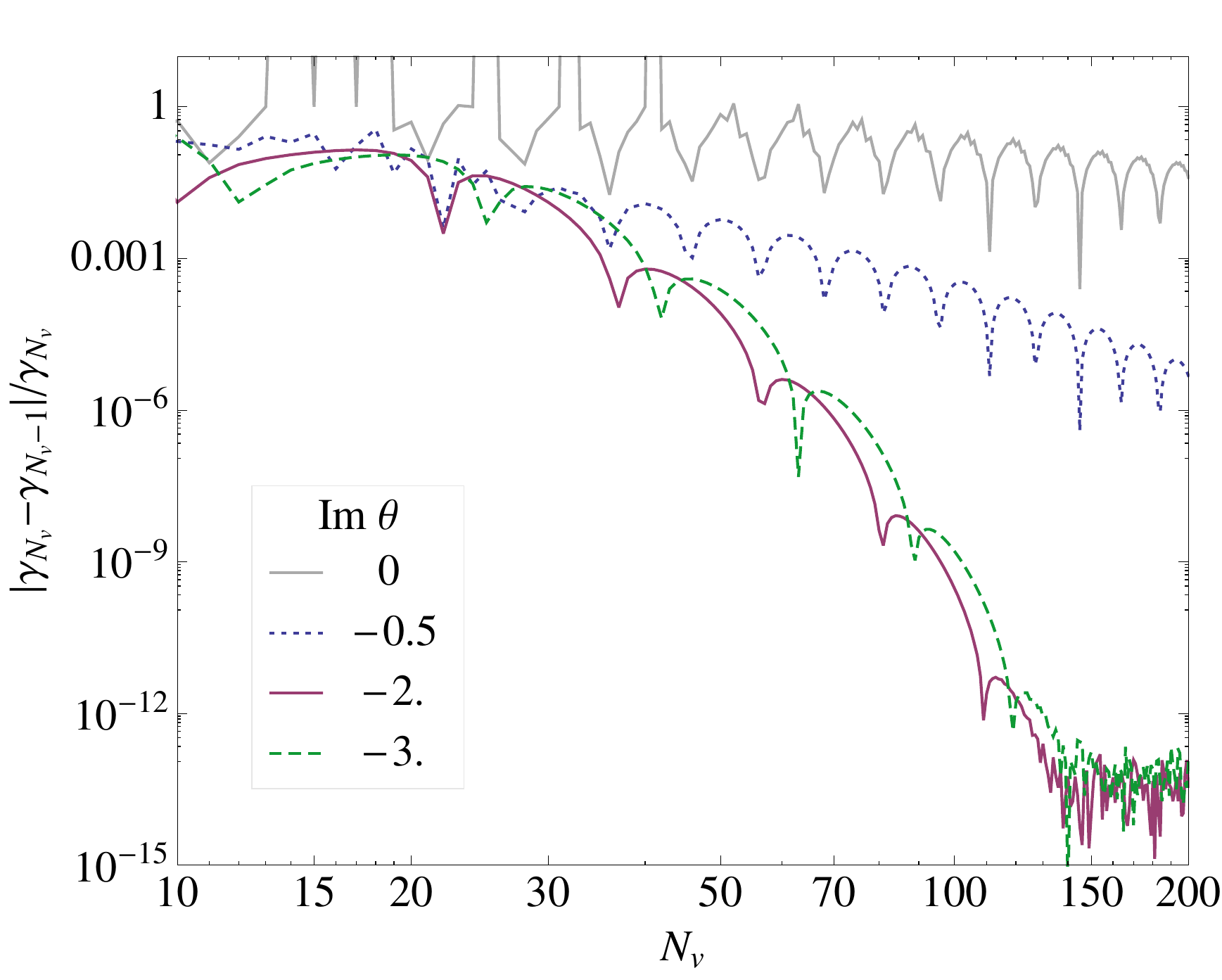}~
   (b)\includegraphics[width=0.3\textwidth,clip=true]
 			  {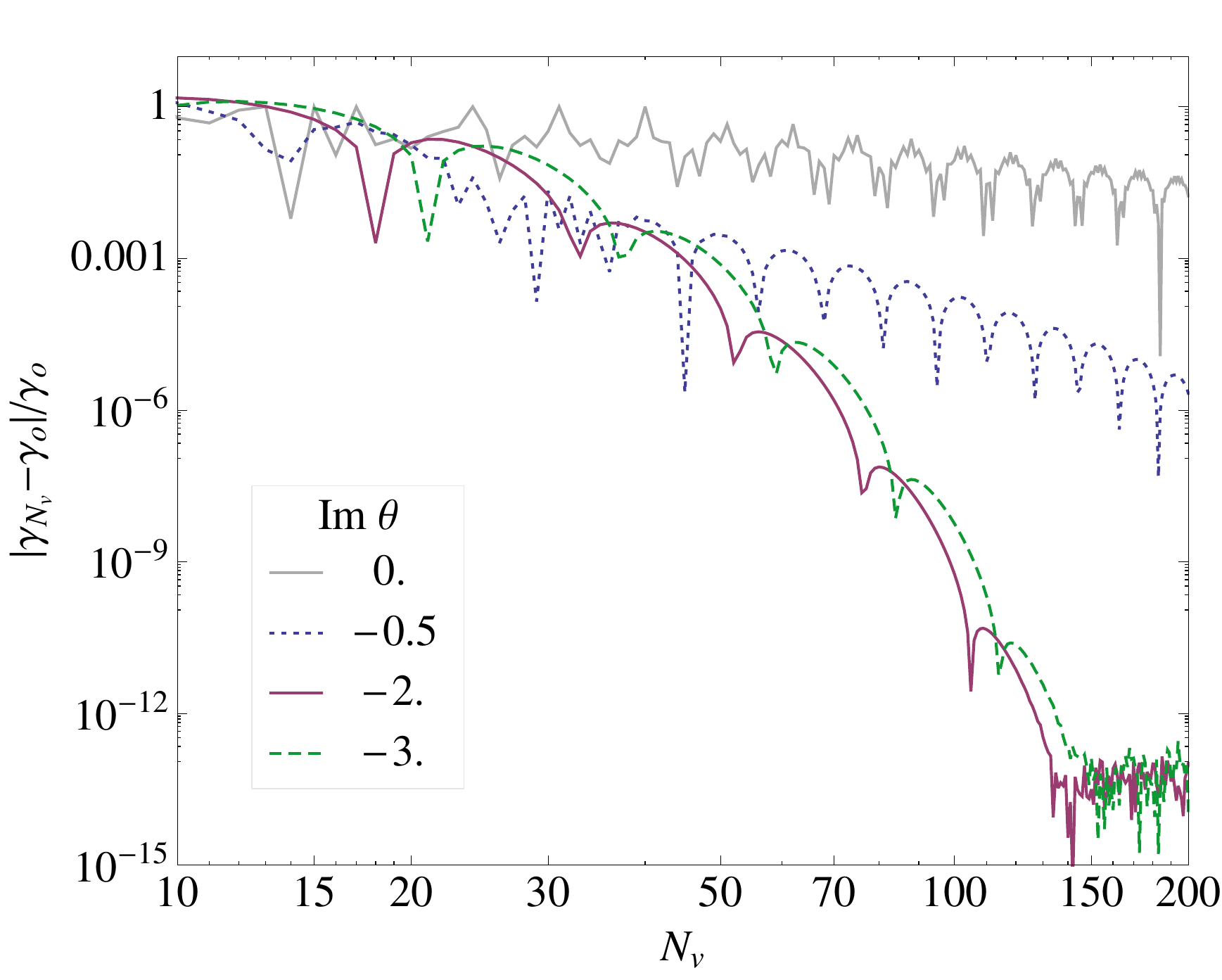}~
  (c)\includegraphics[width=0.3\textwidth,clip=true]{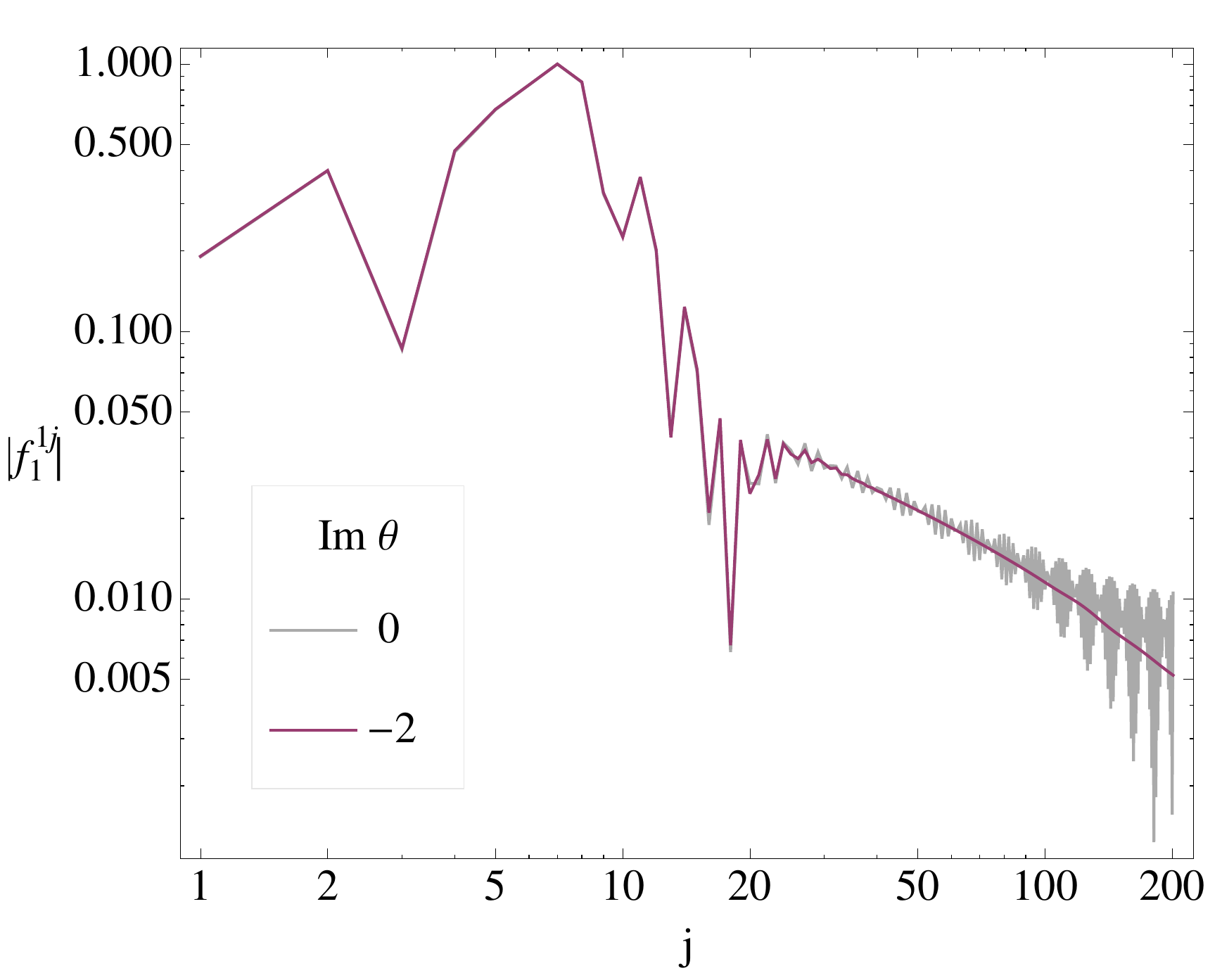}
  \end{center}
  \caption{(Color online)
	(a), relative change of $\gamma_{N_v}$ with $N_v$ and (b), relative
	difference to the exact value from Landau's analysis, 
	for different values of $\Im(\theta)$.
     	(c) Magnitude of Hermite expansion coefficients 
	$|\fp^{1j}|$ of the eigenmode as computed with $\Im(\theta)=0,\,-2$.
	}
\label{f:botConv}
\end{figure*}

The probability distribution function \refeq{eq:bot} can be effectively
approximated with Hermite functions, as indicated by the exponential decay
of the magnitude of Hermite coefficients $\fo^{0j}$ in \reffig{f:bot}(a),
which fall bellow roundoff for $j\gtrsim140$ (with $\vs=2.2$).
Note that both axes are logarithmic in the inset of \reffig{f:bot}(a) 
and exponential convergence corresponds to an envelope of the $\fo^{0j}$ 
with an ever-increasing negative slope.
For $N_v=60$, the residual in $\fo^{0j}$ is of the order
of $10^{-7}$. In \reffig{f:bot}(b) we compare our results for $\gamma$ to
numerical values obtained through Landau's analysis. 
Without spectral deformation, $\Im(\theta)=0$ and $N_v=59$ and $60$, 
agreement is good for small wavelengths, but deteriorates as $\ko$ is increased.
In particular, the mere addition of one Hermite term in the series changes
the results significantly, and for $N_v$ odd\rf{grant67}
the unstable $\ko$ range is narrower than predicted by Landau's analysis. 
Beyond the cutoff wavelength, the damping rate would have to be determined
as in \refsect{s:landau}, but we will not get into this, as we are primarily
interested in growing modes.

To study how the accuracy of our calculations is affected by $N_v$
we fix $\ko=0.36$ and vary the number of Hermite polynomials up to $N_v=200$,
see \reffig{f:bot}(c). As a measure of convergence we plot 
in \reffig{f:botConv}(a)
$|\gamma_{N_v}-\gamma_{N_v-1}|/\gamma_{N_v}$, \ie\ the relative change in
$\gamma$ with the addition of a new term in the Hermite series, while in
\reffig{f:botConv}(b) we compare $\gamma_{N_v}$ with the exact 
value of $\gamma$ derived from Landau's analysis.
Without spectral deformation we observe slow, power-law convergence. Beyond
$N_v=140$ the expansion of $\fo$ only adds numerical noise to the
eigenvalue computation; 
however the relative change in eigenvalues 
is of the order of $10^{-2}$, with pronounced
oscillations with odd and even order truncation in $N_v$, 
see also \reffig{f:bot}(c). 

This even-odd order oscillation behavior has been observed by other 
authors in similar\rf{grant67} or more general settings\rf{camporeale06}.
While some (see Ref.\rf{paskauskas09}) overcome such difficulties 
by averaging the eigenvalues computed over different values of $N_v$, 
such an approach does not warrant convergence and cannot justify the choice
of an expansion over Hermite polynomials, as opposed to an estimation with
a low order method, for instance with finite differences.

Equation \refeq{eq:Vlasov-Poisson-1d-lin-F} shows that the relative
importance of the advection term, responsible for the poor convergence of 
the method, increases with $\ko$.~For relatively
large $\ko$'s, spectral deformation is called for, and with $\theta=-2\ii$ 
and $N_v=60$, agreement with Landau's analysis is recovered
for all $\ko$'s in \reffig{f:bot}(b). 
The unstable range in $\ko$ is now accuratelly retrieved, and
the stability threshold is crossed smoothly,
since damped modes are represented as true eigemodes of  \VlasovLrzSD\ rather
than through the interference of neutral modes.

A closer look at the convergence rate for different values of $\Im(\theta)$ in
\reffig{f:botConv}(a)
reveals that for small values of $-\Im(\theta)$
the convergence still obeys a power-law, yet steeper than the one 
for $\Im(\theta)=0$. 
For large enough values of $-\Im(\theta)$, convergence becomes exponential and
there appears to be no practical advantage in
further increase of $-\Im(\theta)$, since the convergence rate remains
practically the same. 
From the results plotted in \reffig{f:bot}(c), we could say that 
the eigenvalue computation
follows an overdamped oscillation pattern, with very fast relaxation 
towards the exact value. As the expansion coefficients $\fo^{0j}$ 
fall below roundoff at $j\simeq140$, further precision gain with
an increase in $N_v$ ceases, see \reffig{f:botConv}(a) and (b).

It is interesting at this point to examine how rapidly the
expansion coefficients $\fp^{1j}$ of the computed eigenfunction fall off.
With either $\Im(\theta)=0$ or $\Im(\theta)<0$, coefficients $\fp^{1j}$
fall off as a power-low, rather than exponentially as for $\fo^{0j}$, 
see \reffig{f:botConv}(c).
This should have been anticipated: owing to the action of the
advection term the eigenfunctions span all velocity scales, up to the
filamention scale. We therefore need non-vanishing contributions from higher
order Hermite functions to capture such thin scale effects in the eigenfunction. 
The payback of using spectral deformation is a (relatively) accurate 
computation of high $j$ spectral coefficients. 
On the contrary, with $\Im(\theta)=0$ higher $j$
spectral coefficients are subject to strong even-odd order oscillations, see
\reffig{f:botConv}(c).

\subsection{Stability of BGK waves\label{s:BGK}}

\begin{figure*}
  \begin{center}
   	(a)\includegraphics[clip=true]{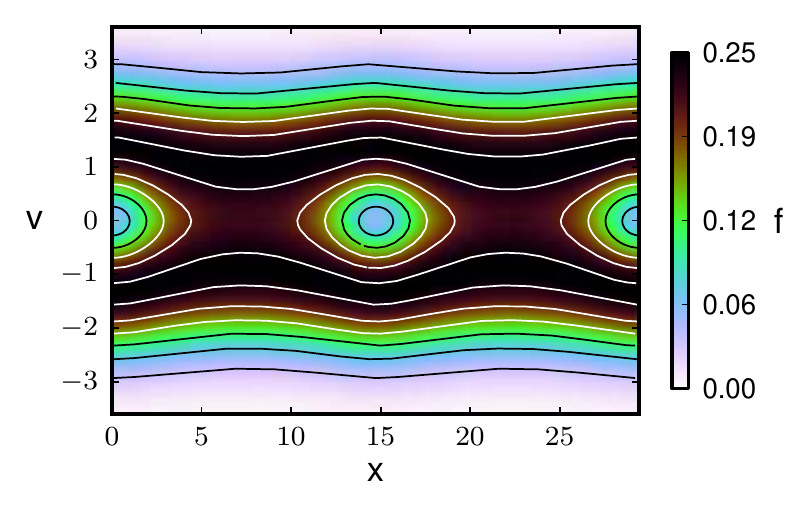}
 	(b)\includegraphics[clip=true]{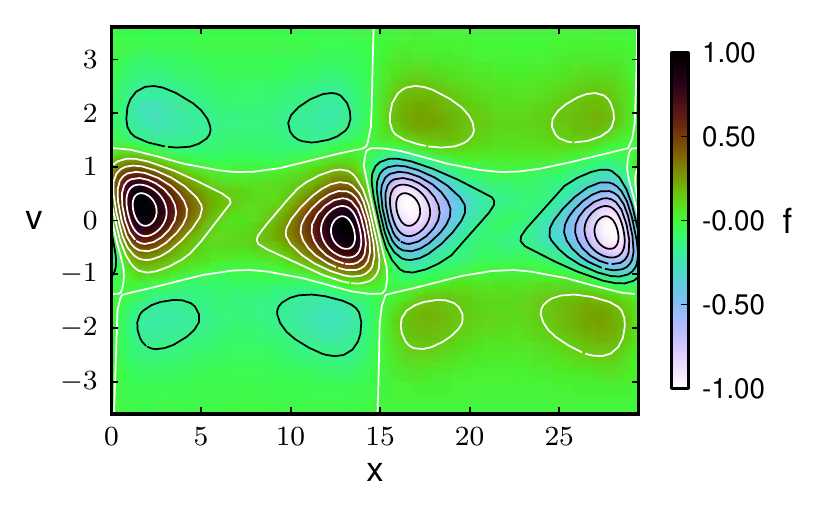}
 	(c)\includegraphics[clip=true]{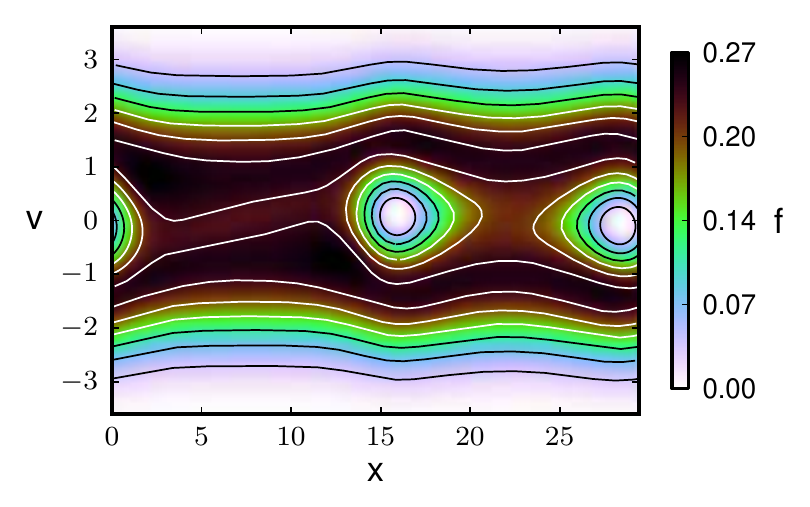}
	(d)\includegraphics[clip=true]{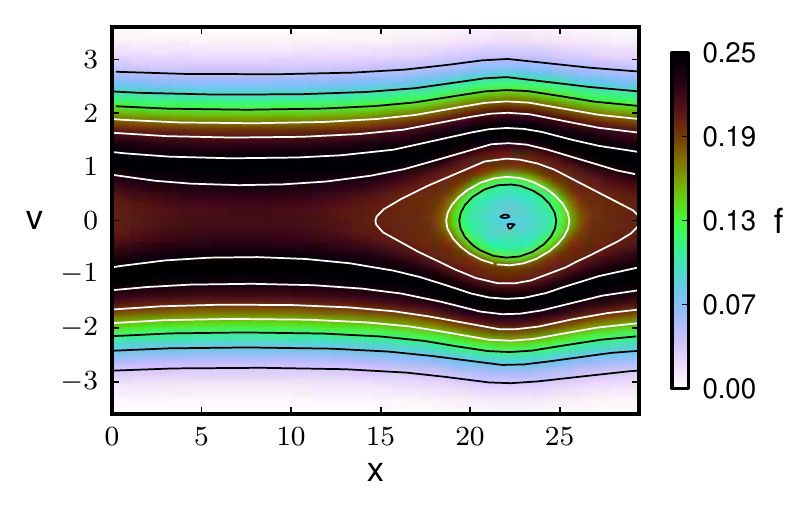}
  \end{center}
  \caption{ (Color online)
	(a) Two spatial periods of the distribution function 
	\refeq{eq:ghizzoHole} with $\mu=0.92,\,\xi=0.90$ (solid lines
	indicate isocontours), 
	(b) the unstable eigenfunction of the $2$-cell equilibrium,
	determined using $N_x=48,\,N_v=110,\,\vs=1.8,\,\theta=-\ii$,
	(c) linear evolution in the neighborhood of the BGK mode, 
	showing the holes approaching each other and triggering fusion,
	(d) late time ($t\simeq1600$) state of the initially 
	perturbed equilibrium obtained 
	with the Eulerian Vlasov code \vador\rf{vador}.
	}
\label{f:ghizzoHole}
\end{figure*}

Bernstein-Greene-Kruskal (BGK) modes\rf{bernstein57} are
stationary, nonlinear electrostatic waves accounting for the presence
of both trapped and untrapped particles. They have attracted a lot of interest
because of their resemblance with the saturated state of nonlinear processes in
plasmas. 
We will study the stability of a particular BGK equilibrium introduced by
Ghizzo \etal\rf{ghizzo1988} to model processes that involve vortex fusion.
Although  no claim
can be made regarding the generality of the shape of $\fo$ prescribed 
in \refref{ghizzo1988}, it serves well to demonstrate how 
our stability calculation can be used to predict vortex fusion.

The chosen equilibrium distribution function is  
\beq\label{eq:ghizzoHole}
  \fo(H)=\frac{\mu}{\sqrt{2\pi}}\frac{2-2\xi}{3-2\xi}\left(1+\frac{H}{1-\xi}\right)e^{-H}\,,
\eeq
where $H(x,v)=v^2/2+\Phi(x)$ is the total energy, $\mu\leq1$ a parameter 
that controls inhomogeneity ($\mu=1$ corresponds to the homogeneous case), 
and $\xi<1$ a parameter that controls the depth of the distribution function's ``depression'' 
or ``hole'' at $x=0$ (see \refref{ghizzo1988} for details). Using Poisson equation, 
one then finds that the potential $\Phi(x)$ 
solves
\beq\label{eq:ghizzoPoisson}
  \Phi''(x)=-\mu\frac{3-2\xi+2\Phi(x)}{3-2\xi}e^{-\Phi(x)}+1\,.
\eeq

In order to allow for subharmonic perturbations,
Ghizzo \etal\ study in Ref.\rf{ghizzo1988} the stability of the equilibrium 
(\ref{eq:ghizzoHole}) for a physical system whose periodicity is $N$ times 
the period, $\Lambda$, of the BGK mode. This system may be viewed as 
an ``$N$-cell replica'' of the basic cell. 
Using the marginal stability analysis developped in\rf{santini70}, 
Ghizzo \etal\ show that when $N\geq2$ the equilibrium is unstable. 
This theoretical prediction is then tested against long-time numerical 
integrations of the Vlasov-Poisson system, with an initial state consisting 
of the equilibrium (\ref{eq:ghizzoHole}) perturbed by a small 
amplitude monochromatic wave whose wavelength is $L=N\Lambda$. 
For $N=1$, the numerical results 
suggest that the equilibrium is stable. For $N\geq2$, instability 
is clearly demonstrated in \refref{ghizzo1988},
as the perturbed equilibrium evolves towards a different final state through
hole merging. 

Here, we compute the unstable modes and their growth rates for the BGK 
equilibrium (\ref{eq:ghizzoHole}) using our Galerkin projection method.~Unless 
otherwise noted, for all Fourier-Hermite calculations 
that follow, we use 
$N_x=24$ points/cell, 
$\vs=1.8$, $N_v$ in the range $40-110$, and $\theta=0$ or $-\ii$.
Since growth rates are not computed
in \refref{ghizzo1988}, we compare our results with the growth rates 
obtained numerically from the resolution of the Vlasov-Possion system 
with the 1D Eulerian Vlasov code \vador\rf{vador}.

\begin{figure}
  \begin{center}
	(a)\includegraphics[width=0.4\textwidth,clip=true]
	    {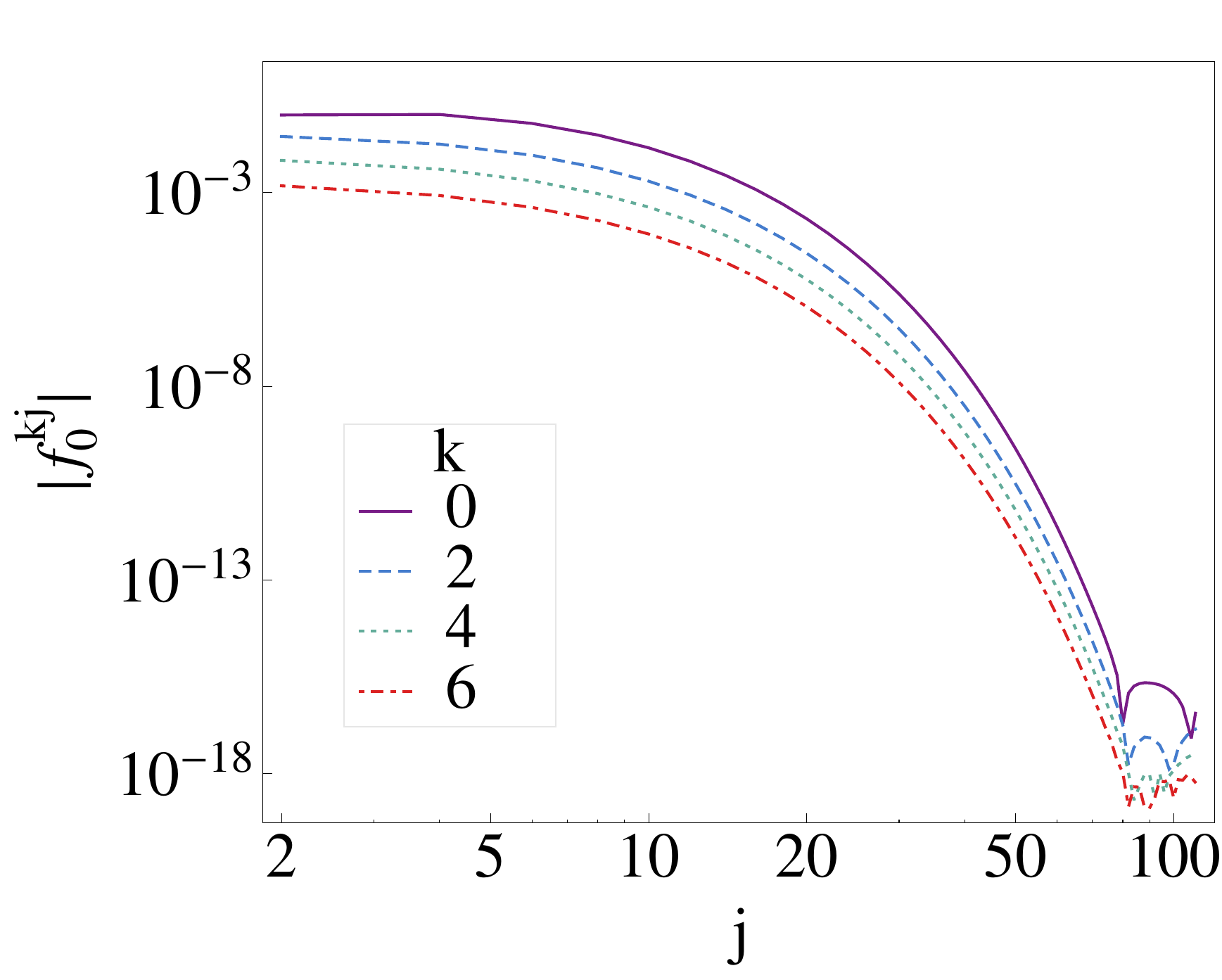}
 	~~~~(b)\includegraphics[width=0.4\textwidth,clip=true]
 	    {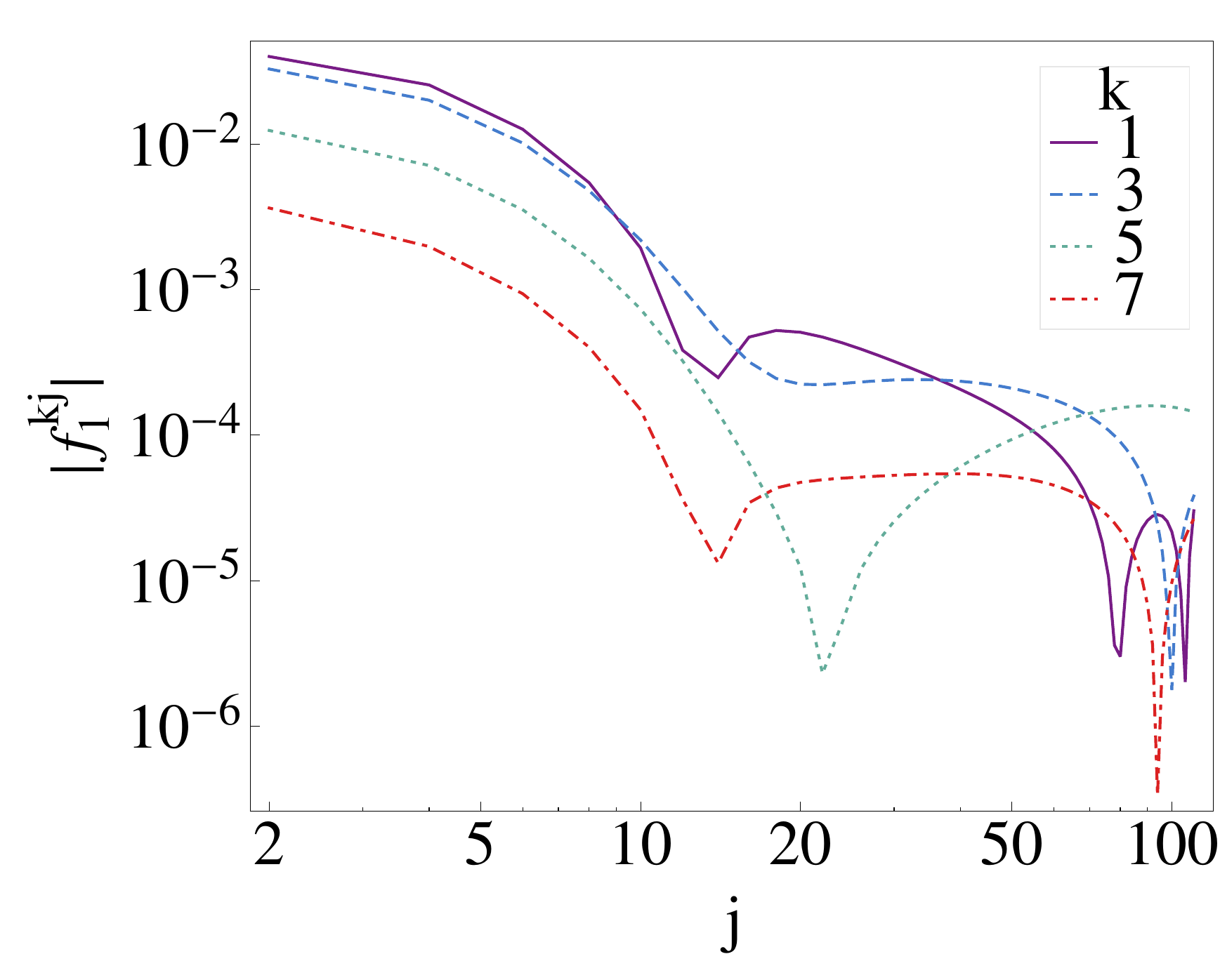}
  \end{center}
\caption{ (Color online)
 	(a) Magnitude of expansion coefficients $|\fo^{kj}|$ of the BGK 
	  mode distribution function~\refeq{eq:ghizzoHole} for $N=2$,
 	computed with
	$v_s=1.8$, $N_x=48$.
 	(b) Magnitude of expansion coefficients $|\fp^{kj}|$ of the corresponding 
	unstable eigenfunction.	
	Only the leading Fourier modes are shown.
}\label{f:ghizzoN2coef}
\end{figure}

Following \refref{ghizzo1988}, we solve Eq.~\refeq{eq:ghizzoPoisson} 
numerically for $\mu=0.92,\,\xi=0.90$. 
The spatial period of the solution is fixed to 
$\Lambda=14.7106$ by specifying the initial condition
$\Phi(0)=\Phi'(0)=0$. Phase space portraits of the distribution function are
characterized by one vortex or ``hole'' for each spatial period, 
as plotted in \reffig{f:ghizzoHole}. Fourier-Hermite expansion coefficients
of $\fo$ fall off exponentially, see \reffig{f:ghizzoN2coef}(a).

\begin{figure}
  \begin{center}
   	(a)\includegraphics[width=0.45\textwidth,clip=true]
	    {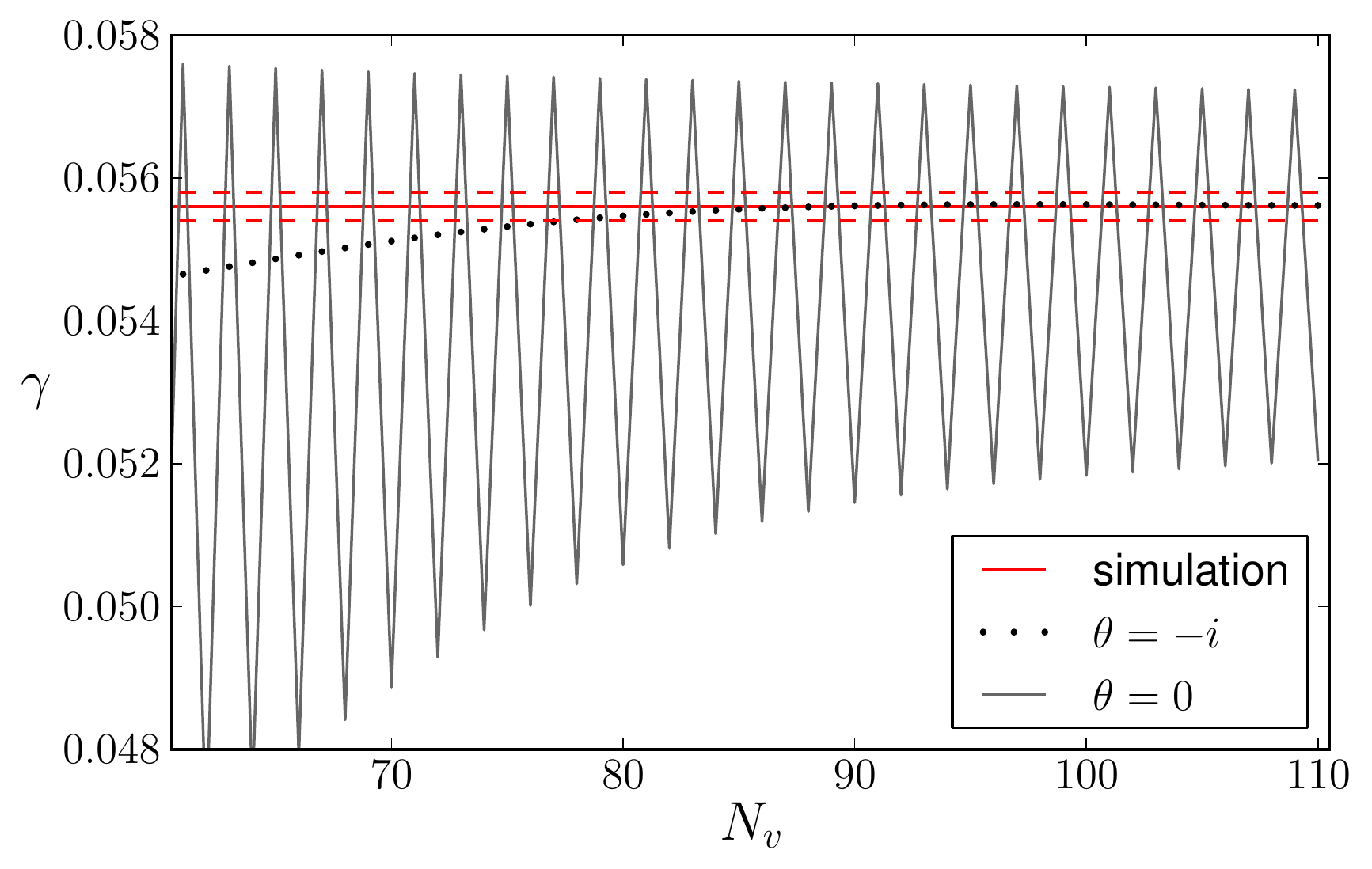}  
 	(b)\includegraphics[width=0.45\textwidth,clip=true]
 	    {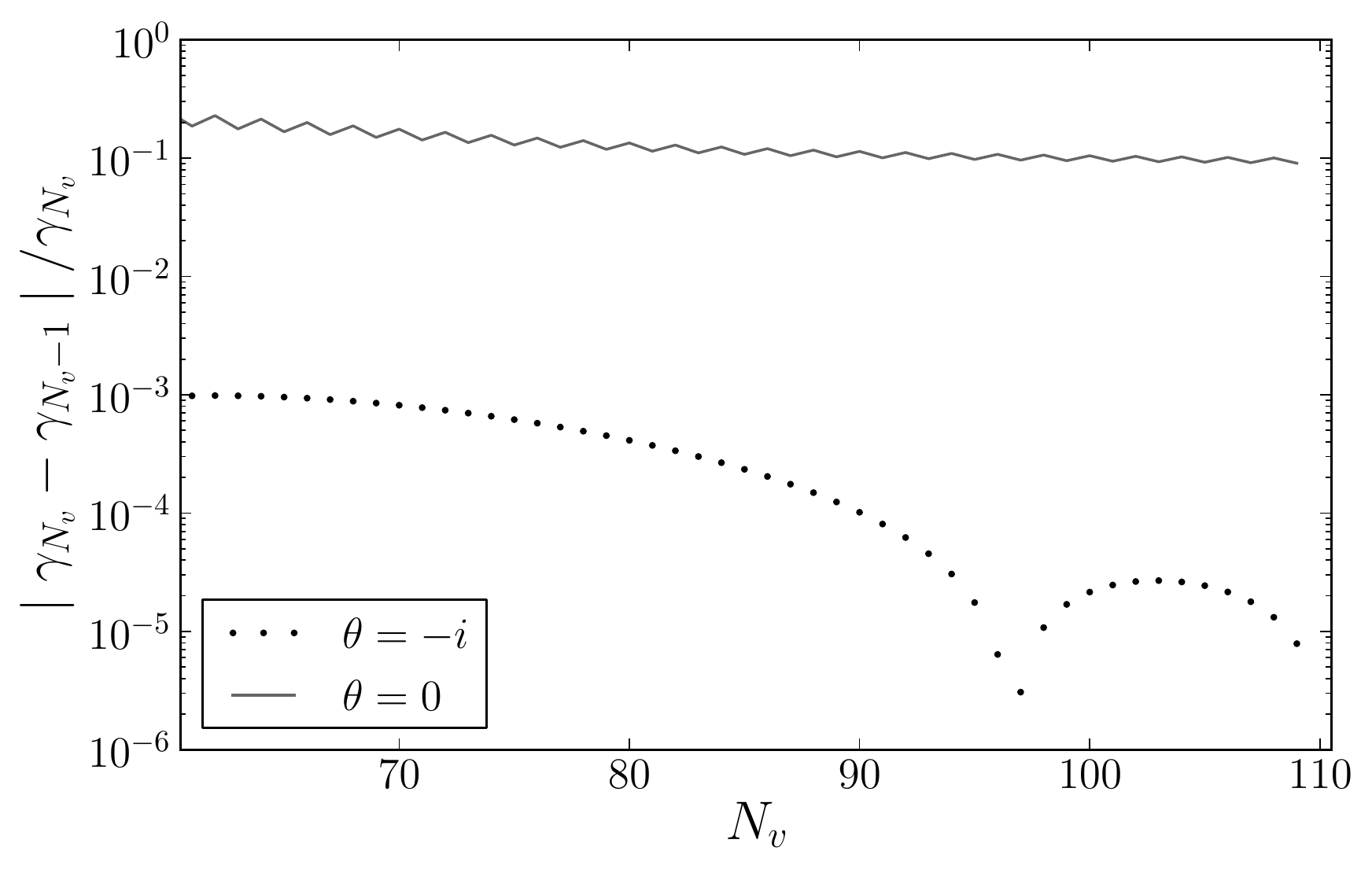} 
  \end{center}
  \caption{ (Color online)
	(a) $N=2$-cell system growth rate $\gamma$ computed
	through Fourier-Hermite expansion with $N_x=48,\vs=1.8,~\theta=0$, $\theta=-\ii$
	and estimated from simulations with the Vlasov code \vador. The red dashed lines
	represent the error bars in the growth rate estimated with \vador.
	(b) Relative rate of change of $\gamma$ with $N_v$. 
		}
\label{f:ghizzoHoleN2comp}
\end{figure}

For $N=2$, the Fourier-Hermite method converges rapidly to the 
growth rate $\gamma=0.05562\pm0.00001$ when we use spectral
deformation with $\theta=-\ii$, with the relative
error falling bellow $10^{-4}$ for $N_v\geq90$, see \reffig{f:ghizzoHoleN2comp}.
The $\theta=0$ computation on the other hand suffers from
large amplitude oscillations between even and odd order in $N_v$
and the convergence rate is slow.

The unstable eigenfunction is plotted in \reffig{f:ghizzoHole}(b).
In \reffig{f:ghizzoHole}(c) we present the predicted 
linear evolution in the neighborhood of the BGK mode,  
showing that the instability causes the holes to approach
each other. This triggers the eventual merging of the two holes,  
illustrated in \reffig{f:ghizzoHole}(d), 
under the nonlinear dynamics.

\begin{figure*}
  \begin{center}
	(a)\includegraphics[width=0.3\textwidth,clip=true]{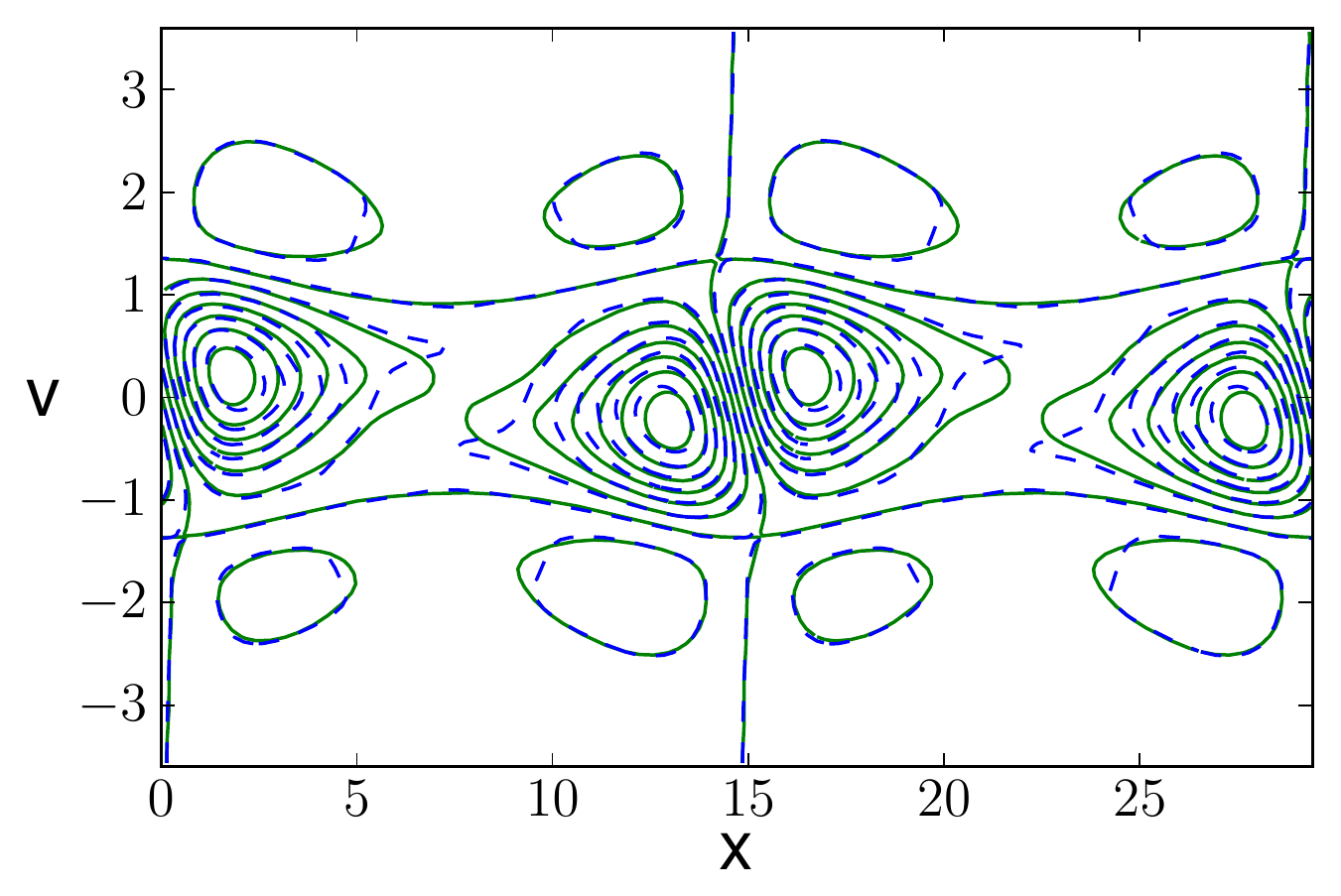}
	(b)\includegraphics[width=0.3\textwidth,clip=true]{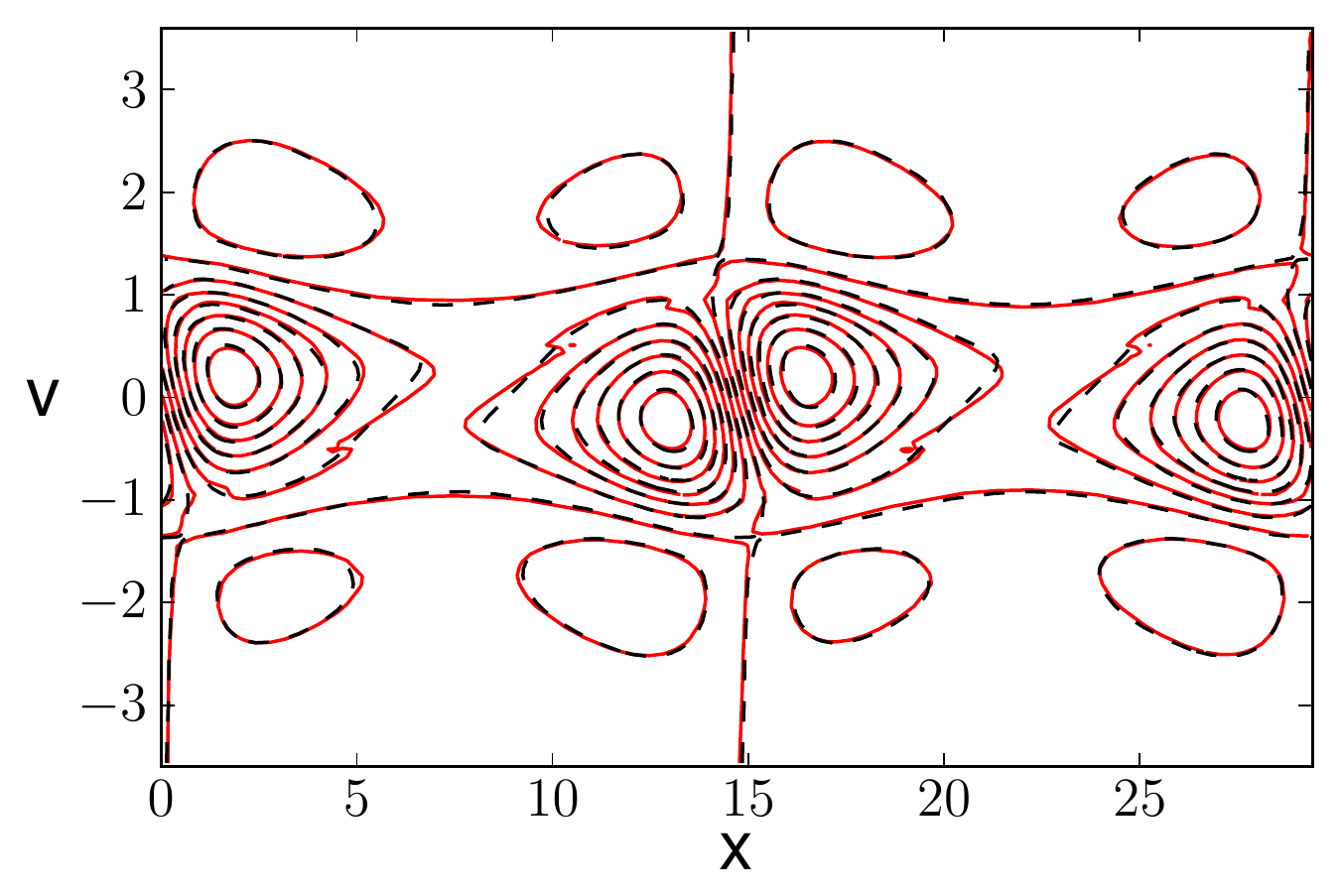}
 	(c)\includegraphics[width=0.3\textwidth,clip=true]{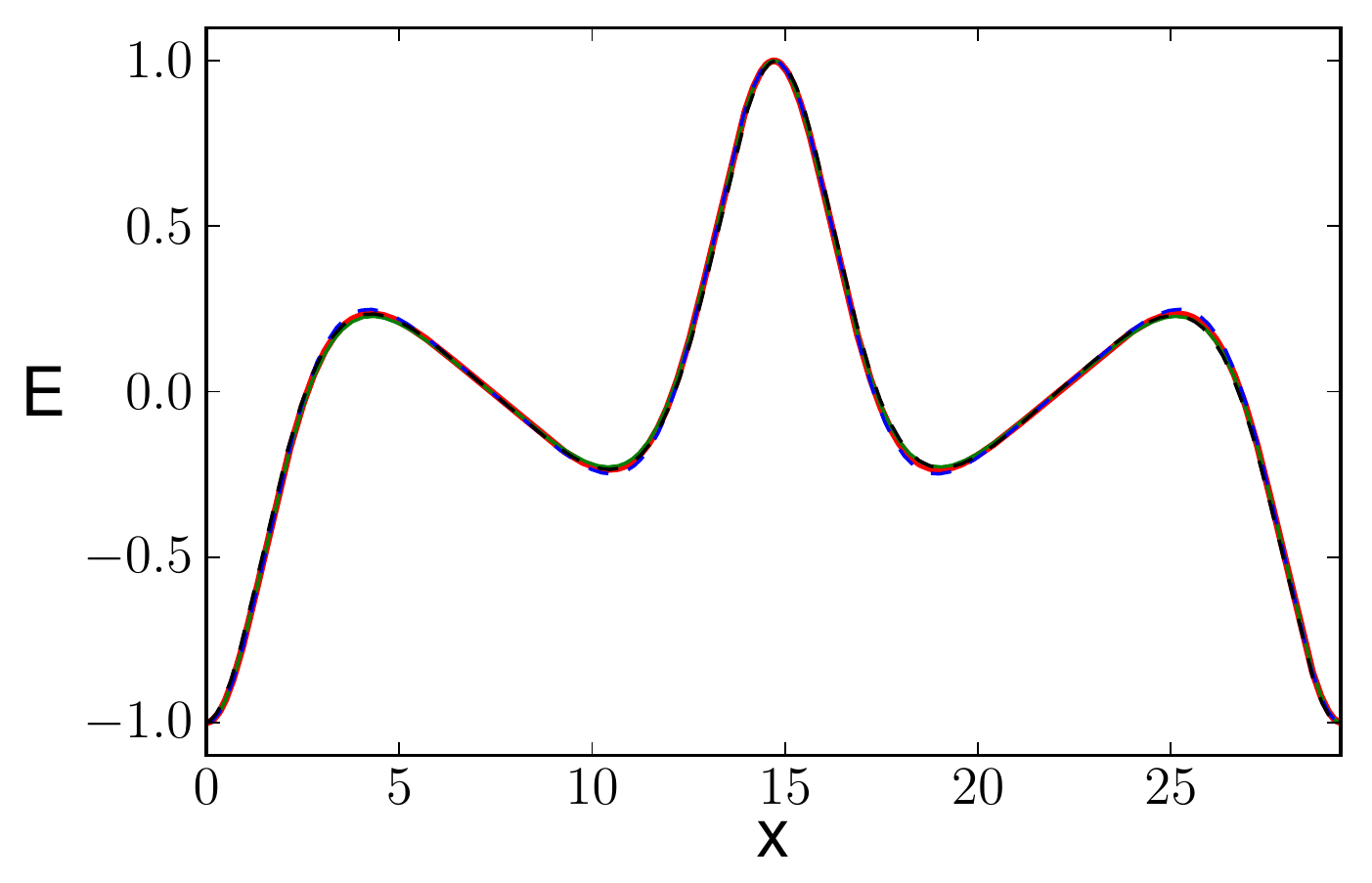}
  \end{center}
  \caption{ (Color online)
	Comparison of the isocontours of the $2$-cell unstable eigenfunction 
	as computed by 
	(a) Fourier-Hermite method with $N_x=48,\vs=1.8$,
	$N_v=109$, $\theta=0$ (blue, dashed line),
	$N_v=110$, $\theta=0$ (green, solid line),
	(b)  Fourier-Hermite method with $N_x=48$, $\vs=1.8$, 
	$N_v=110$, $\theta=-\ii$ (black, dashed line) 
	and numerical integration of a small sinusoidal perturbation 
	with the code \vador\ with $n_x=1920$ and $n_v=1000$ (red, solid line). 
	(c) The corresponding electric field eigenmode, 
	using the same color-code as in Panels (a) and (b).
		}
\label{f:ghizzoHoleN2comp2}
\end{figure*}

The distribution function corresponding to the unstable mode
determined by using the Fourier-Hermite expansion with $\theta=0$ and for
$N_v=109$ and $110$ is shown in \reffig{f:ghizzoHoleN2comp2}(a). 
We can see that the even/odd oscillations in \reffig{f:ghizzoHoleN2comp}(a)
translate into rather large deviations in the shape of the eigenfunctions.

Our results for $\gamma$ and the unstable mode, denoted as $\ev{1}$, 
are compared against those obtained through Vlasov simulations.
We initialize the code \vador\ 
with a perturbed distribution function of the form
$\fo(x,v)\left[1+\alpha\cos(\ko x+\pi/4)\right]$, 
with $\alpha=10^{-5}$, and
follow the evolution of the system in the neighborhood of $\fo$
until $\fp(x,v,t)\equiv f(x,v,t)-\fo(x,v)$ assumes a constant shape, plotted in 
\reffig{f:ghizzoHoleN2comp2}(b), 
and only changes in norm.
We estimate a growth rate $\gamma_{num}=0.0556\pm0.0002$
from the rate of change of the $L^2$ distance $\|f(x,v,t)-\fo(x,v)\|_2$.
Note that the resolution used in \vador, $n_x=1920$, $n_v=1000$,
is much higher than what we needed in the Galerkin method,
while the precision is lower, due to the error in graphically estimating
the growth rate of the perturbations.
We compare the distribution functions corresponding to the predicted 
unstable mode with $\theta=-\ii$,
to the one determined by numerical integration of a small sinusoidal 
perturbation with code \vador\ in \reffig{f:ghizzoHoleN2comp2}(b).
We observe that the two profiles agree rather well,
except for some small-scale structure not captured by our method.
However, the resolution is too different in the two methods 
for a direct comparison at this scale to be meaningfull.

In \reffig{f:ghizzoN2coef}(b), we observe
that the high-order Hermite function components of the unstable eigenmode
do not fall off rapidly, indicating that the eigenfunctions we approximate 
involve fine velocity scales. The utility of the combination 
of spectral deformation and Galerkin projection 
is that it allows the accurate representation of the thermal scale 
effects of the instability, independently of filamentation scale effects.
By contrast, not using spectral deformation ($\theta=0$) results in a much 
coarser approximation of the eigenmodes, with large fluctuations between 
odd and even $N_v$, see \reffig{f:ghizzoHoleN2comp2}(a). 
However, both $\theta=-\ii$ and $\theta=0$ calculations provide
very accurate approximation of the electric field, 
\reffig{f:ghizzoHoleN2comp2}(c), 
as the differences in the distribution function are smoothed out when one 
considers its lower moments.

\begin{figure}
  \begin{center}
   	(a)\includegraphics[width=0.45\textwidth,clip=true]{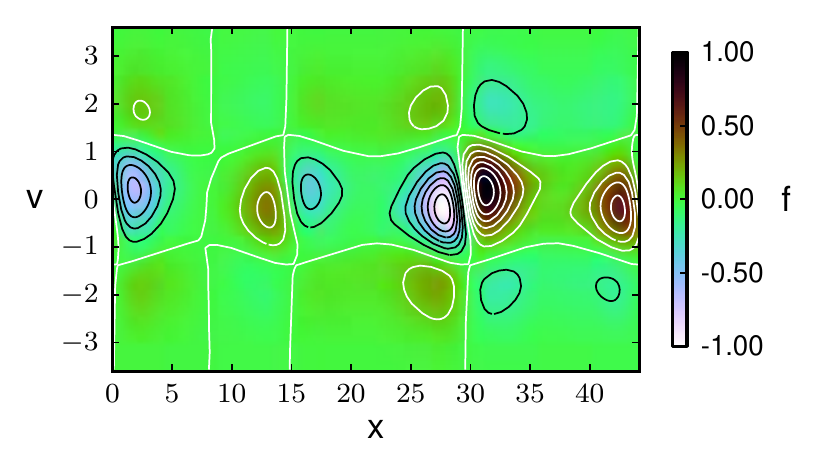}
 	(b)\includegraphics[width=0.45\textwidth,clip=true]{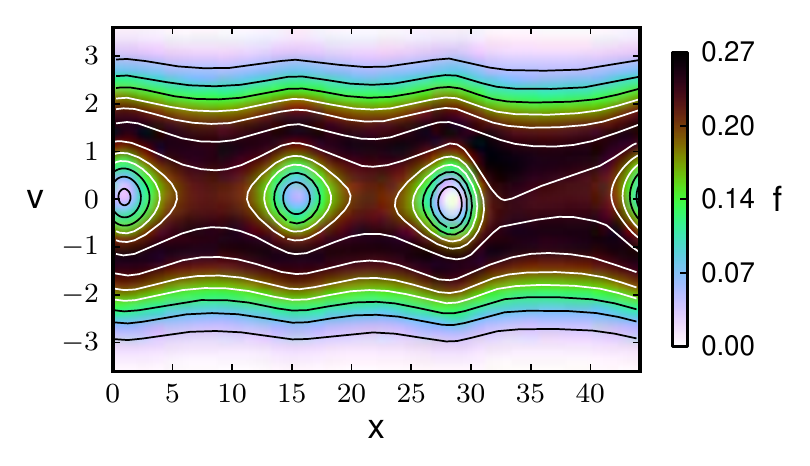}
  \end{center}
  \caption{ (color online)
	(a) The unstable eigenmode of the $3$-cell equilibrium, 
	determined using $N_x=72,\,N_v=110,\,\vs=1.8$, $\theta=-\ii$.
	(b) Predicted evolution in the linear neighborhood 
	of the $3$-cell equilibrium.
	}
\label{f:ghizzoN3}
\end{figure}

The $3$-cell system has a slightly smaller positive eigenvalue $\gamma=0.04856\pm0.00001$, 
with an eigenmode that leads to two-hole fusion, as shown in \reffig{f:ghizzoN3}. 
The numerical simulations in \refref{ghizzo1988} show that this is indeed the case, 
with the third hole subsequently merged with the other two, leading
to an asymptotic one-hole state. We will not present here any detailed
convergence study or comparisons with numerical simulations, as the results
are qualitatively similar to the $2$-cell case.

\begin{figure*}
  \begin{center}
   	(a)\includegraphics[width=0.3\textwidth,clip=true]{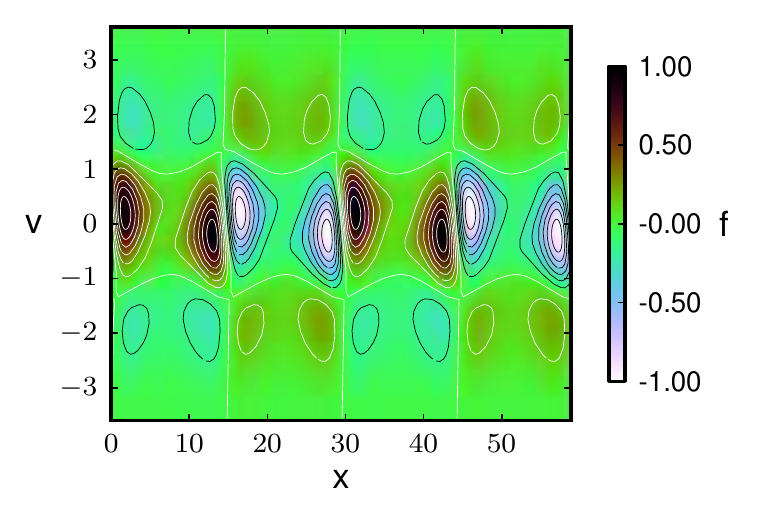}~
     	(b)\includegraphics[width=0.3\textwidth,clip=true]{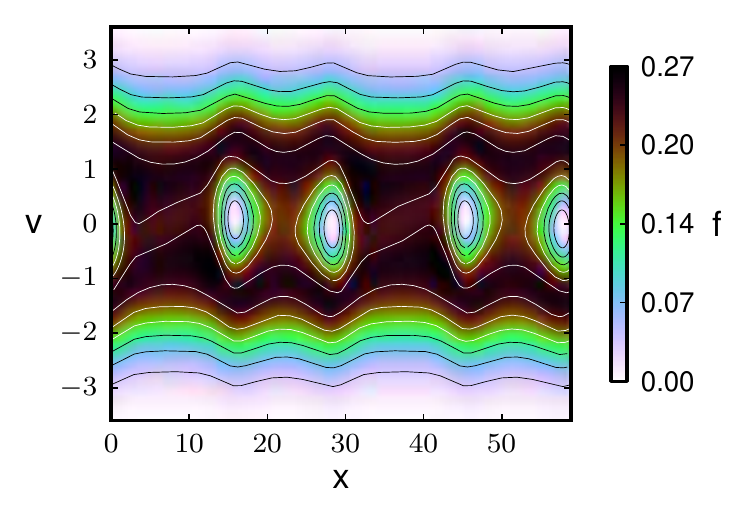}~
   	(c)\includegraphics[width=0.3\textwidth,clip=true]{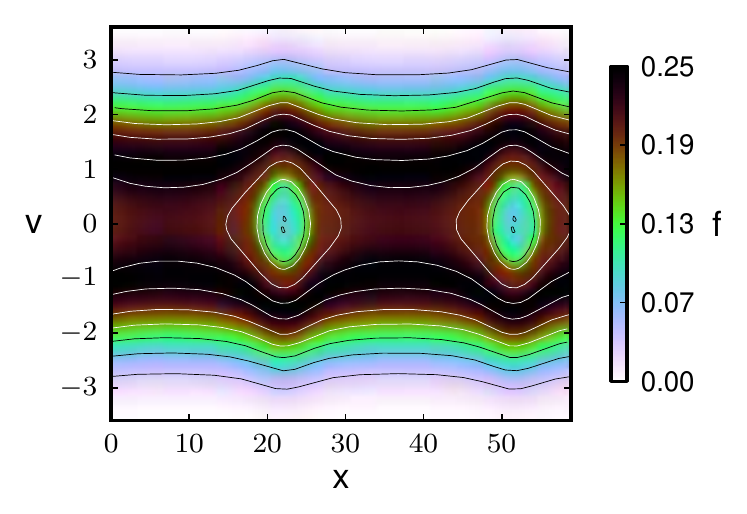}\\
  	(d)\includegraphics[width=0.3\textwidth,clip=true]{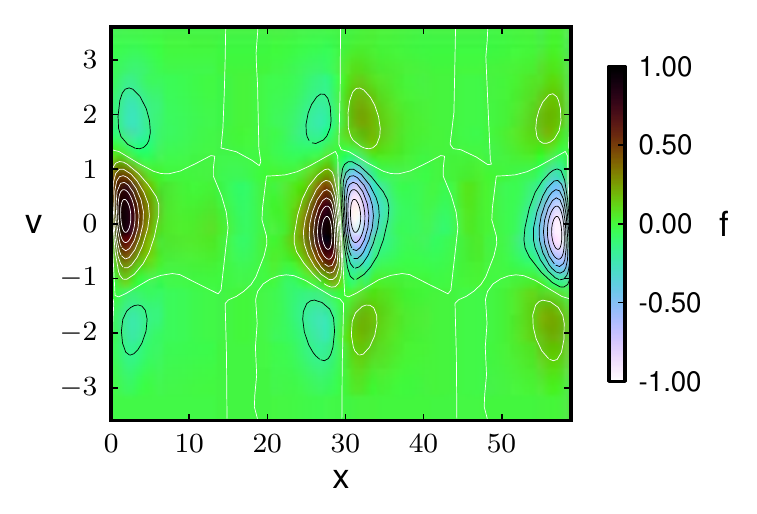}~
  	(e)\includegraphics[width=0.3\textwidth,clip=true]{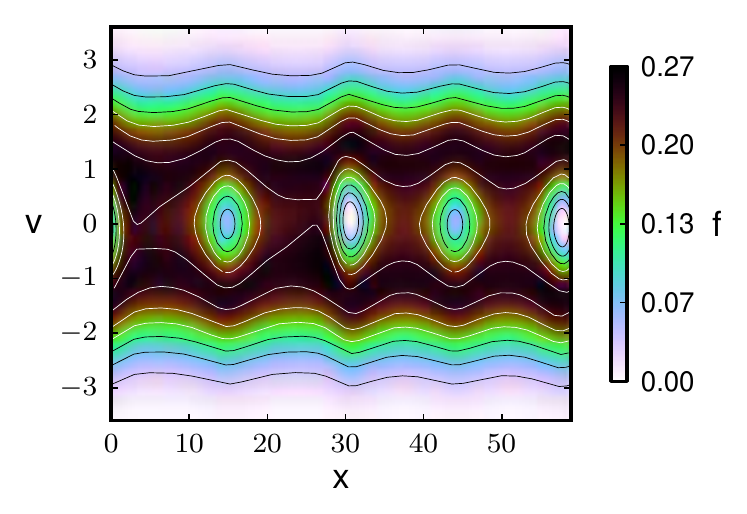}~
  	(f)\includegraphics[width=0.3\textwidth,clip=true]{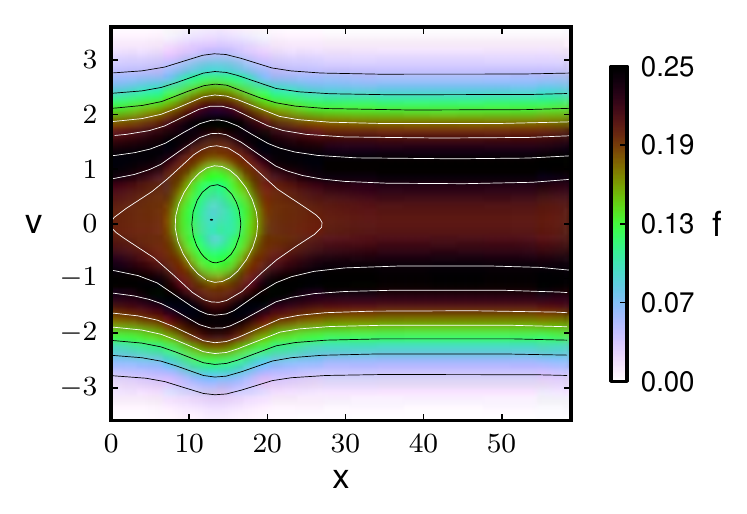}
  \end{center}
  \caption{(color online)
	(a) The most unstable eigenmode $\ev{1}$ of the $4$-cell equilibrium, 
	determined using $N_x=96,\,N_v=90,\,\vs=1.8,\,\theta=-\ii$.
	(b) Linear evolution of a perturbation initialized along $\ev{1}$ and
	(c) final state ($t\simeq1600$) reached after hole 
	merging (using code \vador).
	(d) The second unstable eigenmode $\ev{2}$ of the $4$-cell equilibrium.
	(e) Linear evolution of a perturbation initialized along $\ev{2}$ and
	(f) final state ($t\simeq1600$) reached after hole 
	merging (using code \vador).
	}
\label{f:ghizzoN4}
\end{figure*}

For the $4$-cell state a new possibility arises, since we can think of it as  
two $2$-cell systems stacked together. Thus, the $2$-cell, $k=\kappa_0/2$, 
unstable mode, denoted as $\ev{1}$, is still present, 
with eigenvalue $\gamma_1=0.05562$. 
There is yet another unstable mode, \ev{2}, 
with smaller eigenvalue $\gamma_2=0.04009$,
which only exists for perturbations of wavelength $k=\kappa_0/4$.
The fastest growing mode, $\ev{1}$ is expected to prevail in the case of
a broadband initial fluctuation spectrum.
However it is possible to excite each mode independently, 
by imposing perturbations
of appropriate wavelength. Excitation of $\ev{1}$ 
triggers binary fusion 
of neighboring holes, leading to a final 
state with two holes shown in \reffig{f:ghizzoN4}(a)-(c) which, 
owing to the periodicity of the system, is just the same final state as that 
obtained when $N=2$. By contrast, excitation of $\ev{2}$ 
leads to a more asymmetric hole-fusion senario, with subsequent nonlinear
evolution leading to a one-hole state, see \reffig{f:ghizzoN4}(d)-(e).

\section{Conclusions\label{s:conclusions}}

We have shown that the combination of spectral deformation and Fourier-Hermite
expansion can be used to compute the stability of nonlinear Vlasov-Poisson waves
in an efficient and systematic manner. The computation of unstable
eigenfunctions for the BGK mode of \refsect{s:BGK} illuminated the role of
linear instability in triggering the subsequent vortex fusion. The ability to
detect sub-dominant unstable modes permits a direct assesment of the role of
perturbations of different wavelength in the evolution of the system.
Our method will be used in a future paper to test 
the relevance of considering BGK equilibria in order 
to address Raman saturation.

Spectral deformation was introduced  to handle filamentation scale effects 
through damping of the advective term in Vlasov equation. 
This was the key to achieving exponentially fast convergence of the eigenvalue 
computation and justify the choice of spectral methods. In contrast to
estimates of the growth rate based on numerical integration with 
a Vlasov-Poisson solver,  the combination of spectral
deformation and Galerkin projection scales well with the dimension of phase
space as the exponentially convergent series can be truncated very early, 
while still providing reliable results. We therefore find our 
method to be more suitable for extention to stability calculations in dimensions 
higher than $1$-D, than an approach based on direct integration with 
a Vlasov code.
Moreover, there is in principle no obstacle to generalizing the present 
method to the Vlasov-Maxwell system and to different geometries, 
as the choice of basis
and transformed operator \VlasovLrzSD\ can be adjusted to the problem at hand.
Although our emphasis is on manipulating the structure of the linear operator 
$\VlasovLrz$ itself, rather than on a careful choice of basis, we expect that
the method can be further optimized, if needed, 
through the techniques introduced in \refref{camporeale06}.

\appendix

\section{Hermite basis\label{s:hermite}}

The Hermite polynomials $H_n(\vh)$ we use here follow the standarization 
$H_n(\vh)\sim 2^n \vh^n$ for large $\vh$, see \refref{abramowitz64}. Their explicit definition is
\bseq
	\beq
		H_s(\vh)=(-1)^s e^{\vh^2}\frac{d^s}{d\vh^s}e^{-\vh^2}\,.
	\eeq
They satisfy the orthogonality condition
	\beq
		\int_{-\infty}^{+\infty} H_m(\vh)H_n(\vh) e^{-\vh^2}\,d\vh=\delta_{mn}\sqrt{\pi}2^n n!\,,
	\eeq
\eseq
and therefore
\beq
		\int_{-\infty}^{+\infty} \Psi^m(\vh)\Psi_n(\vh)\,du=\delta_{mn}\,.
\eeq

We note the following relations\rf{abramowitz64}:
\bseq\label{eq:HermiteRec}
	\begin{align}
	    H'_n(\vh)	&= 2n H_{n-1}(\vh)\,,\\
	    H_{n+1}(\vh)&= 2u H_n(\vh)-2n H_{n-1}(\vh)\,.
	\end{align}
\eseq 

Using \refeq{eq:HermiteRec} we can show that the orthonormalized 
Fourier-Hermite basis \refeq{eq:HermiteFunc} satisfies
\bseq\label{eq:RelPsi}
\beq\label{eq:DerivPsi}
	\Psi_n'(\vh)=-\sqrt{2(n+1)}\Psi_{n+1}(\vh)\,,
\eeq
\beq\label{eq:RecPsi}
 	\vh\Psi_n(\vh)=\sqrt{\frac{n+1}{2}}\Psi_{n+1}(\vh)+\sqrt{\frac{n}{2}}\Psi_{n-1}(\vh)\,.	
\eeq
\eseq

Using the identity
\beq
  H_j(\vh+c)=\sum_{k=0}^{j}\binom{j}{k}H_k(\vh)(2c)^{j-k}\,,
\eeq
we can show 
\beq\label{eq:innerPsiShift}
    \int_{-\infty}^{+\infty}\,d\vh\Psi^j(\vh)\Psi_n(\vh+c)=
	\left\{ \begin{array}{l l}
		  0\,, 					&	j<n\,,\\
		  1\,, 					&	j=n\,,\\
		\frac{C_j}{C_n}\binom{j}{n}(-2 c)^{j-n}\,,	& 	j>n\,.
		        \end{array}
	\right.
\eeq

\section{Representation of translation operator\label{s:Umatrix}}

The velocity translation operator $\Uop{\theta}$ 
is represented in Fourier-Hermite basis by the matrix elements
\footnotesize
\beq\label{eq:UrepDef}
	\Urep{\theta}{k}{j}{m}{n} =\int_{-L/2}^{L/2}dx \int_{-\infty}^{+\infty}d\vh\, \Phi^k(x) \Psi^j(\vh)\Uop{\theta_m}\Phi_m(x) \Psi_n(\vh)\,.
\eeq
\normalsize
A direct calculation gives, with the help of \refeq{eq:innerPsiShift},
\beq\label{eq:Urep}
  \begin{split}
	\Urep{\theta}{k}{j}{m}{n}
				  &=\delta_{km}\int_{-\infty}^{+\infty}d\vh\, \Psi^j(\vh)\Psi_n(\vh+\theta_m/\vs)\\
				  &=\delta_{km}
					\left\{ \begin{array}{l l}
					      0\,, 								&	j<n\,,\\
					      1\,, 								&	j=n\,,\\
					      \frac{C_j}{C_n}\binom{j}{n}\left(-2\frac{\theta_m}{\vs}\right)^{j-n}\,,	&	j> n\,.\\
					      \end{array}
					\right.
  \end{split}
\eeq 

However, in our computations we only need the action of 
$\Urep{\theta}{k}{j}{m}{n}$ on a vector
$h^{mn}$ representing a function $h(x,\vv)$.
We can then exploit the sparsity of the representation of the generator of 
velocity translations $\partial_\vv$, 
\[
 (\partial_\vv)^{kj}{}_{lm}=-\vs^{-1}\sqrt{2(m+1)}\,\delta_{kl}\,\delta_{j,m+1}\,,
\]
and employ Krylov subspace approximations\rf{trefethen97} to 
$\Uop{\theta}h(x,\vv)=e^{\theta\frac{\partial}{\partial \vv}}h(x,\vv)$, 
to compute $\Urep{\theta}{k}{j}{m}{n}h^{mn}$ in a stable and efficient manner.

\section{Truncation\label{s:FHtrunc}}

In practice the infinite ladder of equations \refeq{eq:Vlasov-1d-lin-FH-g} 
has to be truncated by setting $f^{N_x+1,j}(t)=f^{k,N_v+1}(t)=0$ for some
cutoff values $N_x$ and $N_v$. We present the truncated form of 
\refeq{eq:VlasovLrz-SD-FH} to emphasize the treatment of boundary terms in
Fourier space and to introduce the notations required to discuss the computation
of the expansion coefficients in \refappe{s:FHcoef}. 
For clarity, we only present the $\theta=0$ case here; extension
to the spectrally deformed case is straightforward. 

The truncated Fourier-Hermite expansions read
\bseq\label{eq:FHexp-trunc}
	\beq\label{eq:FHexpf-trunc}
		f(x_j,\vv_m,t)=\sum_{r=-N_x/2+1}^{N_x/2}\sum_{s=0}^{N_v}f^{rs} \Phi_r(x_j) \Psi_s(\vh_m)\,,
	\eeq
	\beq
		E(x_j,t)=\sum_{r=-N_x/2+1}^{N_x/2} E^r \Phi_r(x_j)\,,
	\eeq
\eseq
$j=1,\ldots, N_x\,,\  m=0,\ldots, N_v\,,$ and $\Phi_r(x)=e^{\ii r \ko x}$, 
$\ko=2\pi/L$ and $x_n=n L/N_x$. 
The calculation of coefficients $f^{rs}$ is presented in 
\refappe{s:FHcoef}.

Note that when taking odd derivatives of \refeq{eq:FHexpf-trunc} with
respect to $x_j$ reality of the result is not ensured, since the term 
$-\ii (N_x/2)\,  e^{-\ii (N_x/2) \ko x_j}$ is not included in the sum.
Thus, we have to set the (presumably small) $N_x/2$ term 
in the discrete Fourier transform of odd derivatives equal to zero, 
see Ref.~\cite[Chapter 3]{trefethen2000}.
Then, proceeding as in \refsect{s:expansionFH} with $\theta=0$ 
we get, when $j\geq1$, 
\beq\label{eq:VlasovLrz-FH}
A^{kj}{}_{lm} = K^{kj}{}_{lm}+ L^{kj}{}_{lm}
				+M^{kj}{}_{lm}\,, 
\eeq
while 
\beq
	A^{k0}{}_{lm}\equiv\begin{cases}
				-ik k_0 \delta_{kl}\frac{v_s}{\sqrt{2}}\delta_{1m}, & k\neq N_x/2,\\
				0,	& k= N_x/2\,,
			\end{cases}
\eeq
and we have introduced
\begin{widetext}
 \begin{align}
	K^{kj}{}_{lm} &=\begin{cases}
	 			-\ii k \ko \vs\delta_{kl} \left(\sqrt{\frac{j}{2}}\delta_{j,m+1}+\sqrt{\frac{j+1}{2}}\delta_{j,m-1}\right),  & k\neq N_x/2,\\
				0,	& k= N_x/2\,,
			\end{cases} \\
	L^{kj}{}_{lm} &=\begin{cases}
	 			- \ii\frac{\pi^{1/4}}{\ko} \frac{\sqrt{2j}}{k-l} \fo^{k-l,0}\, \delta_{j,m+1},  & k\neq l, \text{and }  |k-l|<\frac{N_x}{2},\\
				0,	& \text{otherwise}\,,
			\end{cases}\\
 	M^{kj}{}_{lm} &=\begin{cases}
	 			- \ii\frac{\pi^{1/4}}{\ko} \frac{\sqrt{2j}}{l} \delta_{0m}  \fo^{k-l,j-1},  & l\neq 0,\, l\neq\frac{N_x}{2}, \text{ and }  -\frac{N_x}{2}<k-l\leq\frac{N_x}{2},\\
				0,	& \text{otherwise}\,.
			\end{cases}
\end{align}
\end{widetext}

To store $A^{kj}{}_{lm}$ in a matrix $\mathrm{A}_{\beta\gamma}$ we 
		substitute pairs of indices $(k,j)$ and $(l,m)$ with 
		collective indices $\beta=(k-1)(N_v+1)+j$ 
		and $\gamma=(l-1)(N_v+1)+m$, respectively.

\section{Calculation of Fourier-Hermite coefficients\label{s:FHcoef}}

Calculation of Fourier-Hermite coefficients $f^{rs}$ through a discrete analog 
of 
\beq\label{eq:FHcoef}
 \f^{kj}=\int_{0}^{L}\, dx 
	  \int_{-\infty}^{+\infty}\,d\vh\, 
		\f(x,\vs \vh,t) \Phi^k(x) \Psi^j(\vh)\,,
\eeq
where $\Phi^k(x)=\frac{1}{L}e^{-\ii k \ko x}$,
requires some care. Passage from an integral in $x$ to a
discrete transform is readily handled by a standard Fast Fourier Transform
(FFT).
The analog of the discrete Fourier transform for integration that involves an 
exponentially decaying function $h(x)$ is Hermite-Gauss quadrature 
rule\rf{abramowitz64,NRf77}
\[
\int ^{\infty }_{-\infty} h(\vh) d\vh \simeq \sum^{N}_{i=0} w_i h(\vh_i)\,,
\]
where the \emph{Gauss-Hermite weights} are given by
\beq\label{eq:GHweight}
w_i=\frac{1}{(N+1)\left[\Psi_{N}(\vh_i)\right]^2}\,.
\eeq
The sum has to be evaluated at an appropriate set of points $\vh_i$, 
the Gauss-Hermite \emph{quadrature points} or \emph{abscissas}. 
The quadrature points are given by the roots of $H_{N+1}=0$
and are not available in closed form. 
They can be evaluated as the solution to a symmetric, tridiagonal eigenvalue
problem and are not equispaced\rf{shen09}. 
Evaluation of quadrature weights \refeq{eq:GHweight} 
in a stable fashion is feasible 
through the use of Hermite recursion relations \refeq{eq:RecPsi},
rather than by direct use of \refeq{eq:GHweight}, see 
Refs.~\cite[Chapter 5]{NRf77} and \cite[Remark 4.2]{shen09}.

The Fourier-Hermite coefficients are then given by
\beq\label{eq:FHcoef-trunc}
	f^{k j}(t)=\sum_{n=1}^{N_x}\sum_{i=0}^{N} w_i f(x_n,\vs \vh_i,t) 
		      \Phi^k (x_n) \Psi^j(\vh_i)\,,
\eeq
where $k =-N_x/2+1,\ldots, N_x/2$, $\Phi^k(x)=\frac{1}{N_x}e^{-\ii k \ko x}$,
$j=0,\ldots, N$. In most cases the choice $N=N_v$ 
provides a good approximation of $f^{kj}$, see Ref.~\cite[Chapter 4]{boyd2001}. 
In some cases however, one might want to consider using
$N>N_v$, for reasons related to some special features of 
the Hermite collocation grid.  
As already mentioned, the spacing between adjacent 
Hermite collocation points is not constant and thus the grid 
can by rarefied in regions where the distribution function 
has significant features, such as trapping areas. 
Moreover, the average grid spacing decreases only as $1/\sqrt{N}$ with 
increasing $N$, while at the same time the grid span increases as $\sqrt{N}$
(see \cite[Chapter 17]{boyd2001}).
Therefore, we have to increase $N$ significantly to ensure adequate resolution
in areas of interest, while at the same time we are forced to include 
contributions from many large $\vv$ collocation points, 
where the distribution function is practically zero.

We overcome this problem by a non-standard scheme for the computation 
of $f^{kj}$. Distribution functions of interest fall off faster
than any polynomial in $|\vv|$ for large $\vv$. 
We can therefore introduce appropriate velocity cutoff
values $\vhmin<0<\vhmax$ and replace $\sum_{i=0}^{N}$ in \refeq{eq:FHcoef},
with $\sum_{i=n_1}^{n_2}$, where $\vhmin\leq u_{n_1}$ and $u_{n_2}\leq \vhmax$. 
One can determine 
$n_1,\,n_2$ by counting eigenvalues within $[\vhmin,\vhmax]$ 
of the eigenvalue problem to which $H_{N+1}=0$ reduces,
as prescribed in Ref.~\cite[Lecture 30]{trefethen97}.
We can now pick $N\gg N_v$, ensuring adequate
resolution within the region of interest, without  unnecessarily increasing
the truncation order $N_v$. 
One still has to solve the symmetric, tridiagonal
eigenvalue problem for the collocation points of order $N$. However, 
we only need a small subset $\vhmin<\vh<\vhmax$ of the spectrum and
an efficient algorithm based on bisection can be employed\rf{barth67}. 
Appropriate $\vhmin,\,\vhmax$ and $N$ have to be determined 
empirically for the problem at hand. Good starting 
values for $\vhmin$ and $\vhmax$ are $\vh_{0}$ and $\vh_{N_v}$ respectively, 
since Hermite polynomials of order $N_v$ do not oscillate outside 
$(\vh_{0},\vh_{N_v})$.

To transform from Hermite coefficient space back to physical space efficiently,
Clenshaw recurrence~\cite[Chapter 5]{NRf77} can be used following the FFT 
in \refeq{eq:FHexpf-trunc}.

%


\end{document}